\documentclass[referee]{aa}

\usepackage[utf8]{inputenc}
\usepackage{graphicx}

\usepackage{siunitx}
\usepackage{subfigure}
\usepackage{amsmath}
\usepackage{natbib}
\usepackage{mathtools}
\usepackage{epsf}
\usepackage{xcolor}
\usepackage{multirow}
\usepackage{booktabs}

\newcommand{\kms}{km~s$^{-1}$}
\defcitealias{Bally2000}{BOM2000}
\newcommand{\Ha}{H$\alpha$}
\newcommand{\Hb}{H$\beta$}
\newcommand{\chb}{$c(\mathrm{H\beta})$}
\newcommand{\Oi}{[\ion{O}{i}]$\lambda6300$~\AA}
\newcommand{\atan}{\tan^{-1}}
\newcommand{\Msys}{$M_\mathrm{sys}$}
\newcommand{\Msun}{$M_{\odot}$}
\newcommand{\arc}{\ensuremath{^{\prime\prime}}}

\begin{document}

\title{A spectacular jet from the bright 244-440 Orion proplyd} \subtitle{The MUSE NFM view
\thanks{Based on observations obtained with the MUSE spectrograph on the Very Large Telescope on Cerro Paranal (Chile), operated by the European Southern Observatory (ESO). Program ID: 0104.C-0963(A)}}

\author{A. Kirwan\inst{1}
  \and 
  C. F. Manara \inst{2}
  \and 
  E. T. Whelan \inst{1}
  \and
  M. Robberto \inst{3,4}
  \and
  A. F. McLeod \inst{5,6}
  \and
  S. Facchini \inst{7}
  \and
  G. Beccari \inst{2}
  \and
  A. Miotello \inst{2}
  \and 
  P. C. Schneider \inst{8}
  \and
  A. Murphy \inst{1}
  \and
  S. Vicente \inst{9}
}
  
\institute{Maynooth University Department of Experimental Physics, National University of Ireland Maynooth, Maynooth Co. Kildare, Ireland
 \and 
European Southern Observatory, Karl-Schwarzschild-Strasse 2, 85748, Garching bei M{\"u}nchen, Germany
 \and
Johns Hopkins University, 3400 N. Charles Street, Baltimore, MD 21218,USA
 \and
Space Telescope Science Institute, 3700 San Martin Dr, Baltimore, MD 21218, USA
 \and
Centre for Extragalactic Astronomy, Department of Physics, Durham University, South Road,  Durham DH1 3LE, UK
\and 
Institute for Computational Cosmology, Department of Physics, University of Durham, South Road, Durham DH1 3LE, UK
\and 
Dipartimento di Fisica, Universit\'a degli Studi di Milano, via Celoria 16, Milano, Italy
\and
Hamburger Sternwarte, Universit{\"a}t Hamburg, Gojenbergsweg 112, 21029 Hamburg, Germany
\and Instituto de Astrofísica e Ciências do Espaco, Universidade de Lisboa, OAL, Tapada da Ajuda, P-1349-018 Lisboa, Portugal
}

\titlerunning{A spectacular jet from the bright 244-440 Orion proplyd} 
\date{\today}

\abstract{
In this work we present the highest spatial and spectral resolution integral field observations to date of the bipolar jet from the Orion proplyd 244-440 using Multi-Unit Spectroscopic Explorer (MUSE) narrow-field mode (NFM) observations on the Very Large Telescope (VLT). We observed a previously unreported chain of six distinct knots in a roughly S-shaped pattern, and by comparing them with Hubble Space Telescope (HST) images we estimated proper motions in the redshifted knots of {9.5 mas~yr$^{-1}$ with an inclination angle of \ang{73}}, though these quantities could not be measured for the blueshifted lobe. Analysis of the [\ion{Fe}{ii}] and [\ion{Ni}{ii}] lines suggests jet densities on the order of $\sim 10^5~{\rm cm}^{-3}$. We propose that the observed S-shaped morphology originates from a jet launched by a smaller source with $M_\star < 0.2$~\Msun\, in orbital motion around a larger companion of $M_\star \simeq 0.5$~\Msun\, at a separation of $30-40$~au. The measured luminosities of the knots using the [\ion{O}{i}]$\lambda6300$~\AA\, and [\ion{S}{ii}]$\lambda6731$~\AA\, lines were used to estimate a lower limit to the mass-loss rate in the jet of $1.3 \times 10^{-11}$~\Msun~yr$^{-1}$ and an upper limit of $10^{-9}$~\Msun~yr$^{-1}$, which is typical for low-mass driving sources. 
While the brightness asymmetry between the redshifted and blueshifted lobes is consistent with external irradiation, further analysis of the [\ion{Ni}{ii}] and [\ion{Fe}{ii}] lines suggests that photoionization of the jet is not likely to be a dominant factor, and that the emission is dominated by collisional excitation. The dynamical age of the jet compared to the anticipated survival time of the proplyd demonstrates that photoevaporation of the proplyd occurred prior to jet launching, and that this is still an active source. These two points suggest that the envelope of the proplyd may shield the jet from the majority of external radiation, and that photoionization of the proplyd does not appear to impact the ability of a star to launch a jet.}


\maketitle

\section{Introduction}
\label{sec:intro}

\begin{figure}[htbp]
  \centering
  \includegraphics[width=\linewidth, trim={0cm 0cm 0cm 0cm}, clip=true]{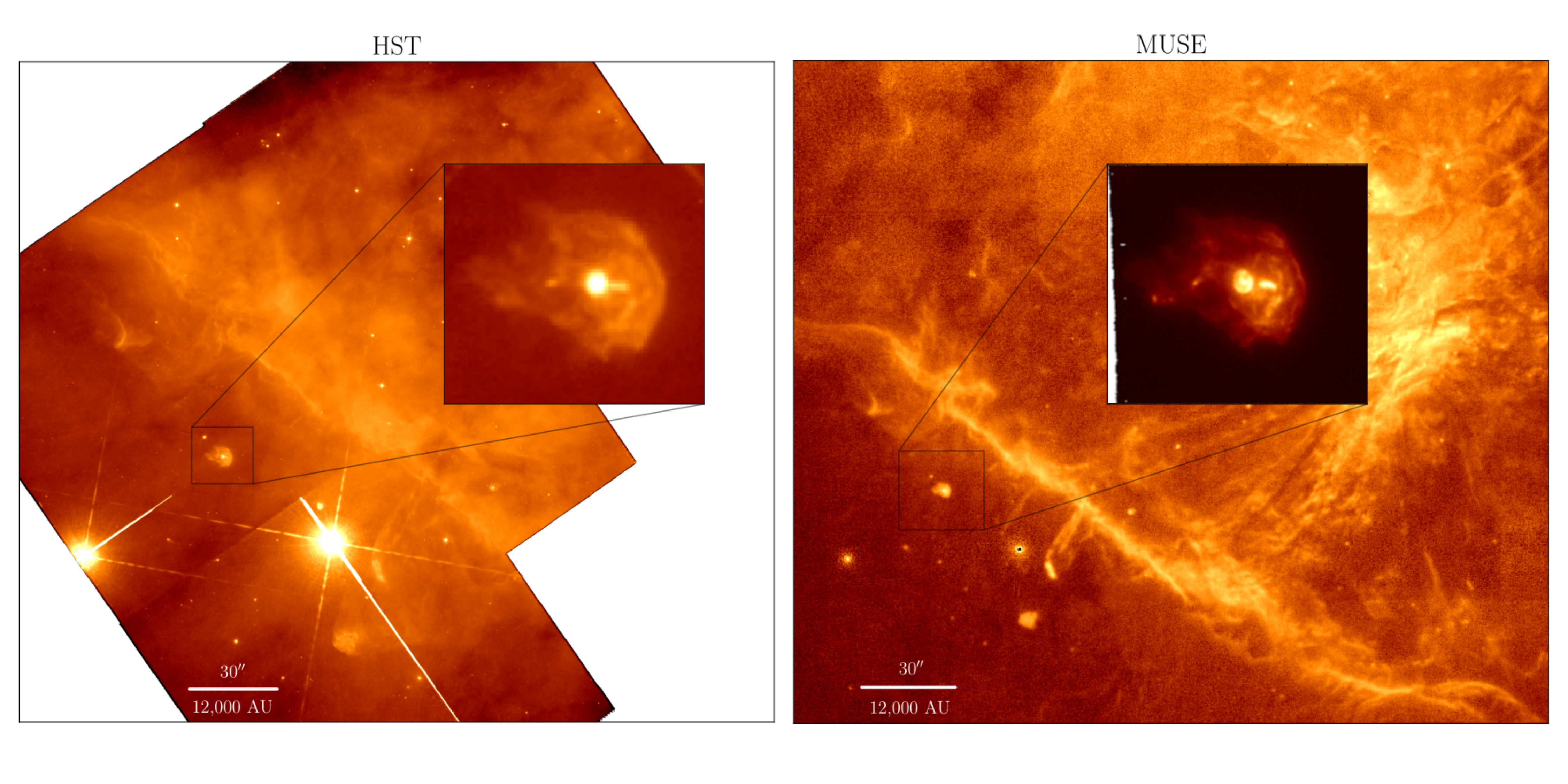}
  \caption{HST (left) and MUSE (right) field images of the region around 244-440. The fields and insets in both images are shown in the \Oi\, emission. The stars $\theta^2$ Orionis A and B are seen just below the proplyd to the southwest and southeast, respectively. The inset in the MUSE panel is a flux-integrated image from our current NFM observations discussed in Section~\ref{sub:vlt_muse}.}%
  \label{fig:muse_hst_insets}
\end{figure}

Astrophysical jets are near-ubiquitous phenomena in the evolution of low- and intermediate-mass young stellar objects~\citep[YSOs, ][]{frank2014}, and have similarly been observed in high-mass YSOs as well~\citep{Marti1993ApJ, McLeod2018Natur}. The formation of clumpy, shock-heated condensations known as Herbig-Haro (HH) objects is linked with ejection events, and as such the study of these objects allows us to better understand the mass-loss history of the star. The morphology and physical conditions of the jets are determined by the characteristics of the ejection mechanism as well as by the surrounding environments: single, isolated stars may launch straight, bi-polar outflows, while driving sources in denser regions may experience side winds from more massive stars, possibly compounded with a high proper motion, that induce a curvature in the jet~\citep{Bally2001ApJ, Raga2009AA}; in binary systems, companions may induce warping in the inner disk, precession, or orbital motion that causes a ``wiggling'' jet morphology~\citep{masciadri2002, Lai, Murphy2021, Erkal2021AAa, Kirwan2022AA}.

The exact mechanism of jet launching is not known, but current research has found a strong correlation between ejection events and mass accretion and circumstellar disk dynamics~\citep{RayNewAR, Whelan2014, nisini2018}. Studying such ejections in a variety of environments is crucial to fully understand the physics behind them. Regions such as the Orion Nebula Cluster (ONC) provide a unique laboratory for studying the early stages of star formation, due to the high stellar density and the relative closeness of the region ($d~\sim$ 400 pc). Moreover, the environment allows for jets to be examined under the most extreme conditions. 

Many YSOs within the ONC have been found to possess externally illuminated, photo-evaporating disks often surrounded by nebular structures and ionization fronts (IFs). With the discovery of these ``proplyds'' (PROtoPLanetarY DiskS) came the observations that many of them are associated with jets and HH objects~\citep{ODell1997AJ, Bally2001ApJ, Ricci2008AJ}. They typically are one-sided and exhibit a C-shaped curvature pointed away from the massive ionizing stars within the cluster. Proplyds in the dense, inner regions of the ONC present us with a lifetime problem as the measured mass-loss rates due to ultraviolet (UV) irradiation are too high and should rapidly evaporate the disk~\citep{Clarke2007MNRAS}. However, near-infrared (NIR) excess is still observed in $\sim 80$\% of the ONC stars, implying that their disks may survive longer than predicted. 
This implies that either initial disk masses are large ($>1$~\Msun) and therefore unstable, or that the massive ionizing stars have ages $\leq 0.1$~Myr, which is short compared to the region average~\citep[$2-4$~Myr, see][]{Beccari2017AA.}. Estimates of disk masses in the inner regions of the ONC have not generally been larger than  $10^{-2}$~\Msun~\citep{Henney1999AJ}, and more recent surveys with the Atacama Large Millimeter/submillimeter Array (ALMA) have indicated a maximum dust mass of $\sim 80~M_{\oplus}$~\citep{Eisner2018ApJ}. 
A combination of age spread and stellar dynamics, with the youngest stars migrating inward to the central regions, may help solve the lifetime discrepancy~\citep{Winter2019MNRAS}.
{This should have direct implications on the morphology of the jets and on the accretion-outflow connection in photoevaporated disks}, granting deeper insights into the mechanics of jet launching and the impact of the environment.

The giant proplyd 244-440, known also as V* V2423 Ori and HH 524, is located at a distance of $400 \pm23$~pc~\citep{Gaia2021b}. With a width of 3\farcs5 ($\sim 1400$~au), it is one of the largest proplyds in the ONC~\citep*[][hereafter BOM2000]{Bally2000}. It is not located in the inner core of the ONC, as its large IF of radius $\sim 2\farcs0$ ($\sim 800$~au) points toward $\theta^2$~Orionis A, located southwest of the proplyd and beyond the prominent photodissociation region (PDR) known as the ``Orion Bar,'' as shown in Figure~\ref{fig:muse_hst_insets}. Hubble Space Telescope (HST) images in \Ha\, were interpreted by~\citet{Bally2001ApJ} as showing a nearly edge-on disk in silhouette with a size of approximately 0\farcs15 $\times$ 0\farcs6, and a tilted jet with a $\sim \ang{15}$ difference between the jet axis and the disk minor axis. \citetalias{Bally2000} further note that the star appears offset from the center of the disk by $\sim$~0\farcs1, and suggest a binary system where one star is hidden within the disk. Spectral observations of the source suggest that it is a low-mass star (< 1~\Msun; see Appendix~\ref{sec:appendixA}). 

\citet{Henney1999AJ} initially proposed a mass-loss rate for {the proplyd} of $\sim 1.5 \times 10^{-6}$~\Msun\, yr$^{-1}$, although~\citet{Winter2019MNRAS} posit that this is an overestimation, suggesting instead a mass-loss rate on the order of $\sim 5 \times 10^{-8}$~\Msun\, yr$^{-1}$. Direct estimates of the disk$+$envelope mass are difficult, but range from as high as 0.01~\Msun\, \citep{Bally1998AJ} based on {millimeter} measurements, to as low as $\sim 5 \times 10^{-3}$~\Msun\, with the VLA \citep{Sheehan2016ApJ}.\footnote{We note that the VLA is free$-$free (FF) emission dominated, and while it is possible that the proplyds are FF dominated as well, this latter value is presented with caution.} While the values of~\citet{Bally1998AJ} suggest evaporation times $t_e \sim 10^4$~yr~\citep[see also][]{Henney1999AJ}, those of~\citet{Sheehan2016ApJ} and~\citet{Winter2019MNRAS} suggest $t_e \sim 0.1-0.2$~Myr, which ultimately set a lower limit on the age of the proplyd.

In this paper, we present the first detailed analysis of the bi-polar jet associated with this proplyd. In Section~\ref{sec:observations} we discuss our MUSE data and data reduction process. In Section~\ref{sec:morph}, we outline a proper motions study and offer an explanation of the complex curvature of the jet. The extinction, physical conditions, and mass-loss rate are estimated in Section~\ref{sec:physical_properties}. We discuss our results in Section~\ref{sec:discussion} and provide a summary in Section~\ref{sec:conclusions}.

\section{Observations \& data reduction}
\label{sec:observations}

\subsection{VLT/MUSE}%
\label{sub:vlt_muse}

Observations of proplyd 244-440 were obtained on 23 October 2019 with the Multi-Unite Spectroscopic Explorer (MUSE) under program ID 104.C-096 (PI: C. F. Manara). The instrument was operated in narrow field mode (NFM) with adaptive optics (AO) under clear sky conditions, which allows sampling of the target with a spatial resolution of 0\farcs025 pixel$^{-1}$. The image quality delivered by the AO system was measured in the data as $\sim0\farcs14$.  
For at least 50\% of the total observation, the coherence time was $> 6$~ms. Data reduction was performed with the MUSE pipeline (v2.8) with ESO Reflex using standard calibrations and recipes~\citep[for a detailed description of the MUSE data reduction pipeline, see][]{Weilbacher2020A&A}. Further details of the data reduction process will be described in a forthcoming paper. A final cube was produced spanning the entire nominal wavelength range of the MUSE instrument ($\sim4750-9350$\AA), with a field of view (FOV) of 8\farcs4 $\times$ 8\farcs6.

This final cube was further processed in Python to remove local continuum and nebular contributions, using a method similar to that described by~\citet{AgraAmboage2009AA}. For each spectral emission region of interest, a subcube spanning $\sim100$\AA\, was extracted from the primary cube and continuum-subtracted by fitting a second-order polynomial to the spectrum at each spaxel in the subcube~\citep[see][for a description of this process and considerations of the wavelength-dependence of the PSF]{Kirwan2022AA}. Additionally, to account for nebular background and foreground line emission we constructed a mean, local background spectrum for each subcube by sampling regions away from the proplyd envelope but close enough to be representative of the emission profile, and subtracted this from the subcube. Extinction was calculated using the Balmer decrement under Case B assumptions, which is further discussed in Section~\ref{sub:extinction}.

\subsection{Archival HST images}%
\label{sub:archival_hst_images}

Two epochs of archival HST Wide-Field Planetary Camera (WFPC2) images were used to estimate the proper motions of the knots (see Section~\ref{sub:knot_id}). The observations were obtained on 14 November 1995 and 17 September 1998 as part of the General Observer programs GO 5976 (PI: J. Bally) and GO 6603 (PI: C. O'Dell), respectively. We have focused on the F631N (\Oi, both epochs) filters in this study, which have total exposure times $t_{\rm exp} = 2100$~s and $t_{\rm exp} = 1200$~s respectively. Additionally, GO 5976 has observations in the F673N ($t_{\rm exp} = 2100$~s) and F791W filters ($t_{\rm exp} = 200$~s), and GO 6603 has observations in the F656N ($t_{\rm exp} = 600$~s) and F814W filters ($t_{\rm exp} = 60$~s). The jet is seen in the F631N and F673N filters, while only the envelope is seen in the F791W and F814W filters. 
A description of the observations can be found in~\citet{ODell1997AJ}.\footnote{The science-ready data described here may be found on the MAST archive at~\url{http://dx.doi.org/10.17909/y66h-8p10}.} 
To increase the S/N, we combined the [\ion{O}{i}] images for each epoch into {two} individual stacks. 
We then separately aligned each stack of images to {its own} common reference frame in Python using the \texttt{mpdaf}\footnote{https://github.com/musevlt/mpdaf} function \texttt{align\_with\_image}, and median-combined the two epochs separately. The jet-free images for each epoch were finally subtracted from the \Oi\, images to better isolate the jet emission, as shown in Figure~\ref{fig:hst_epochs}.

\begin{table}[t]
\centering
\caption{Emission line fluxes}
\label{tab:244_440_fluxes}

\begin{tabular}{rrrrrrrr}
\hline\hline
$\lambda_{\mathrm{air}}$ (\AA) &             Ion & F$_\mathrm{E1}$ & F$_\mathrm{E2}$ & F$_\mathrm{E3}$ & F$_\mathrm{W3}$ & F$_\mathrm{W2}$ & F$_\mathrm{W1}$ \\
\hline
                        5158.8 & [\ion{Fe}{iii}] &          \ldots &          \ldots &          \ldots &          \ldots &           14.87 &           13.23 \\
                        6300.3 &    [\ion{O}{i}] &           70.15 &           26.98 &           73.41 &          160.32 &          183.44 &           99.47 \\
                        6363.8 &    [\ion{O}{i}] &           23.47 &            7.47 &           25.75 &           53.78 &           60.33 &           39.66 \\
                        6716.4 &   [\ion{S}{ii}] &           15.48 &          \ldots &           15.95 &           52.92 &           71.16 &           49.25 \\
                        6730.8 &   [\ion{S}{ii}] &           28.59 &          \ldots &           33.08 &           98.02 &          146.27 &          101.20 \\
                        7155.2 &  [\ion{Fe}{ii}] &           11.88 &            8.38 &           11.03 &           20.39 &           19.78 &           17.19 \\
                        7172.0 &  [\ion{Fe}{ii}] &            2.71 &            2.47 &            2.99 &            6.45 &            5.72 &            4.80 \\
                        7255.8 &  [\ion{Ni}{ii}] &          \ldots &        \ldots   &          \ldots &          \ldots &            7.89 &            4.27 \\
                        7320.0 &   [\ion{O}{ii}] &           71.86 &           31.25 &           46.76 &          \ldots &          \ldots &          \ldots \\
                        7330.2 &   [\ion{O}{ii}] &           54.27 &           23.45 &           36.61 &        \ldots   &          \ldots &          \ldots \\
                        7377.8 &  [\ion{Ni}{ii}] &            8.11 &            4.82 &            5.72 &           12.36 &           12.77 &           12.22 \\
                        7388.2 &  [\ion{Fe}{ii}] &            2.10 &          \ldots &          \ldots &            3.35 &            3.20 &            2.80 \\
                        7411.6 &  [\ion{Ni}{ii}] &            1.29 &          \ldots &          \ldots &          \ldots &            1.68 &            2.04 \\
                        7452.5 &  [\ion{Fe}{ii}] &            2.83 &            2.23 &          \ldots &            5.78 &            5.81 &            5.65 \\
                        8578.7 &  [\ion{Cl}{ii}] &            2.64 &          \ldots &          \ldots &            3.74 &            2.36 &            5.51 \\
                        8617.0 &  [\ion{Fe}{ii}] &           13.64 &            9.27 &           12.27 &           23.93 &           23.47 &           20.48 \\
                        9052.0 &  [\ion{Fe}{ii}] &          \ldots &          \ldots &           \dots &            6.44 &            5.54 &            4.46 \\
\hline\hline
\end{tabular}
\tablefoot{Dereddened mission line fluxes for the knots in the Proplyd 244-440 jet, in units of 10$^{-17}$ erg/s/cm$^{2}$. Line fluxes were computed through the Gaussian fitting of spectral profiles extracted from the aperture sizes {given in Section~\ref{sub:knot_id}.} The average uncertainty in the lines fluxes is $\sim$5 $\times$ 10$^{-18}$ erg/s/cm$^{2}$, which we calculate from the RMS noise in adjacent portions of the spectrum. Empty entries denote positions where either no knot emission is observed or the emission is below the detection threshold, or the knot emission could not be reliably disentangled from the proplyd envelope.}
\end{table}

\section{Morphology}
\label{sec:morph}

\subsection{Knot identification in the MUSE data} 
\label{sub:knot_id}

The angular size of the giant proplyd 244-440 and the spatial sampling of MUSE NFM+AO allow for a detailed view of the bipolar jet, which is detected in multiple emission lines. In many cases, most notably the [\ion{O}{i}] emission, one observes both the jet and envelope. Lines such as [\ion{Fe}{ii}] and [\ion{Ni}{ii}] only show jet emission while others (e.g., [\ion{Ar}{iii}]) show the envelope and the ionization front. By comparing the different morphologies traced by different emission lines, it is possible to clearly identify the jet knots by removing the ``contamination'' by nebular and envelope emission (see Figures~\ref{fig:append_A2}$-$\ref{fig:append_A4}). 

\begin{figure}[htpb]
  \centering
  {\includegraphics[width=0.48\linewidth, trim={0cm 0cm 0cm 0cm}, clip=true]{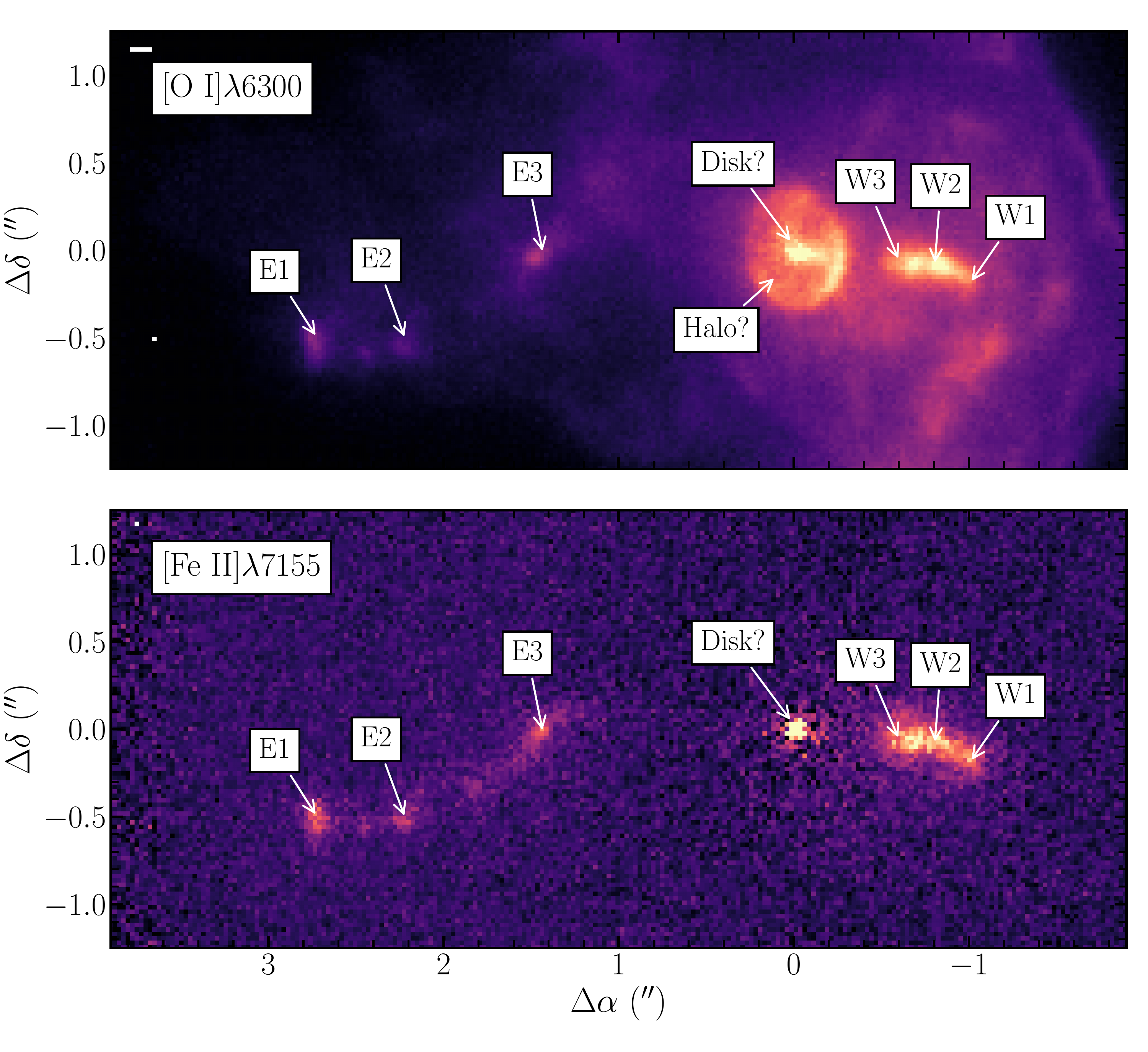}}
  \includegraphics[width=0.48\linewidth, trim={0cm 0cm 0cm 0cm}, clip=true]{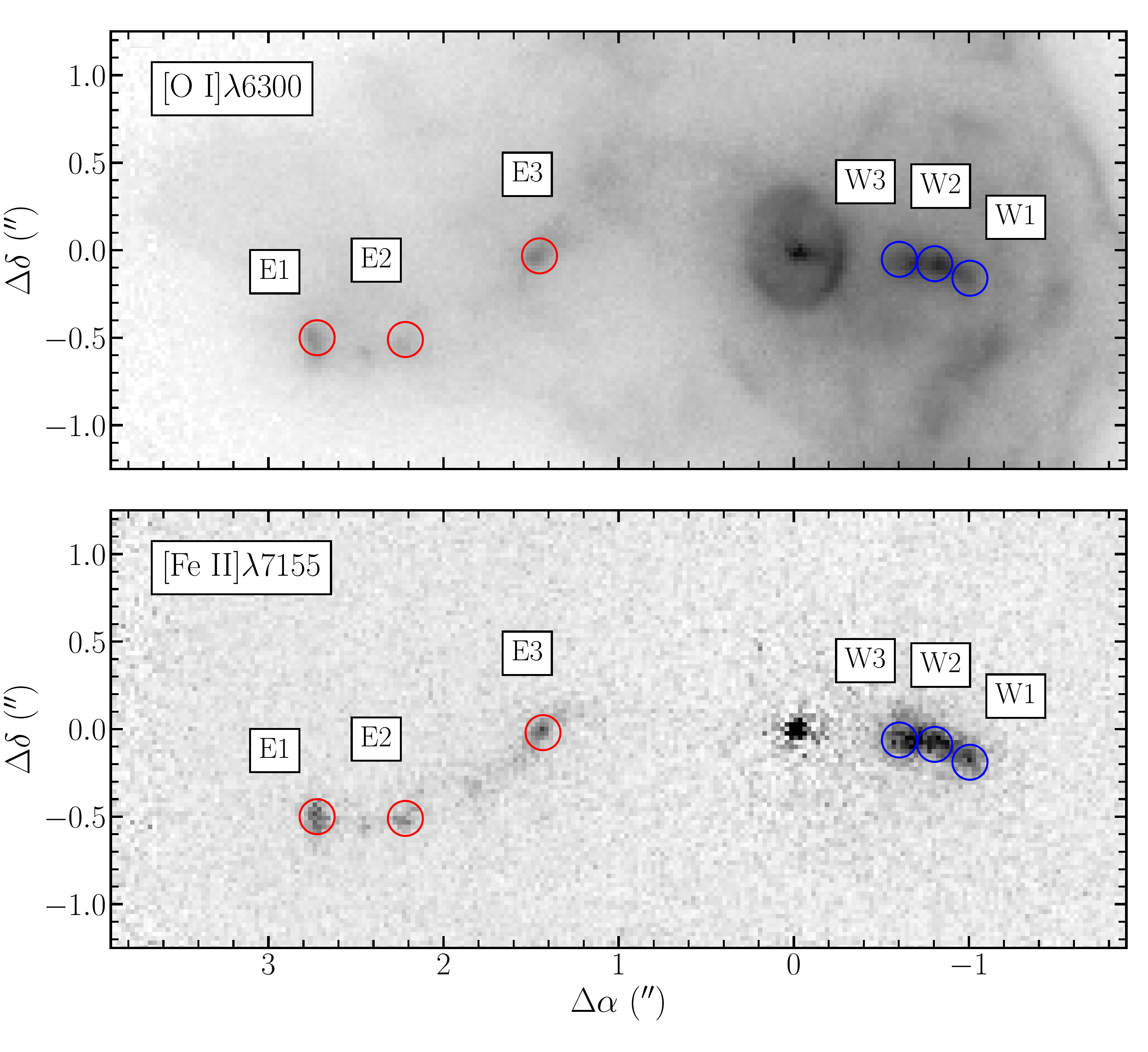}
  \caption{Identified knots in the MUSE images. Left: The [\ion{O}{i}]$\lambda6300$ (top) and [\ion{Fe}{II}]$\lambda7155$ (bottom) emission lines with the tentative disk and halo noted in the [\ion{O}{ii}] panel. Emission is seen very close to the source, but whether this is the disk or the jet is undetermined. {Right}: The same images in grayscale, with the redshifted and blueshifted knots indicated by colored circles.}%
  \label{fig:244_440_annotated}
\end{figure}

\begin{figure}[htpb]
   \centering
   \includegraphics[width=\linewidth, trim={0cm 0.5cm 0cm 0cm}, clip=true]{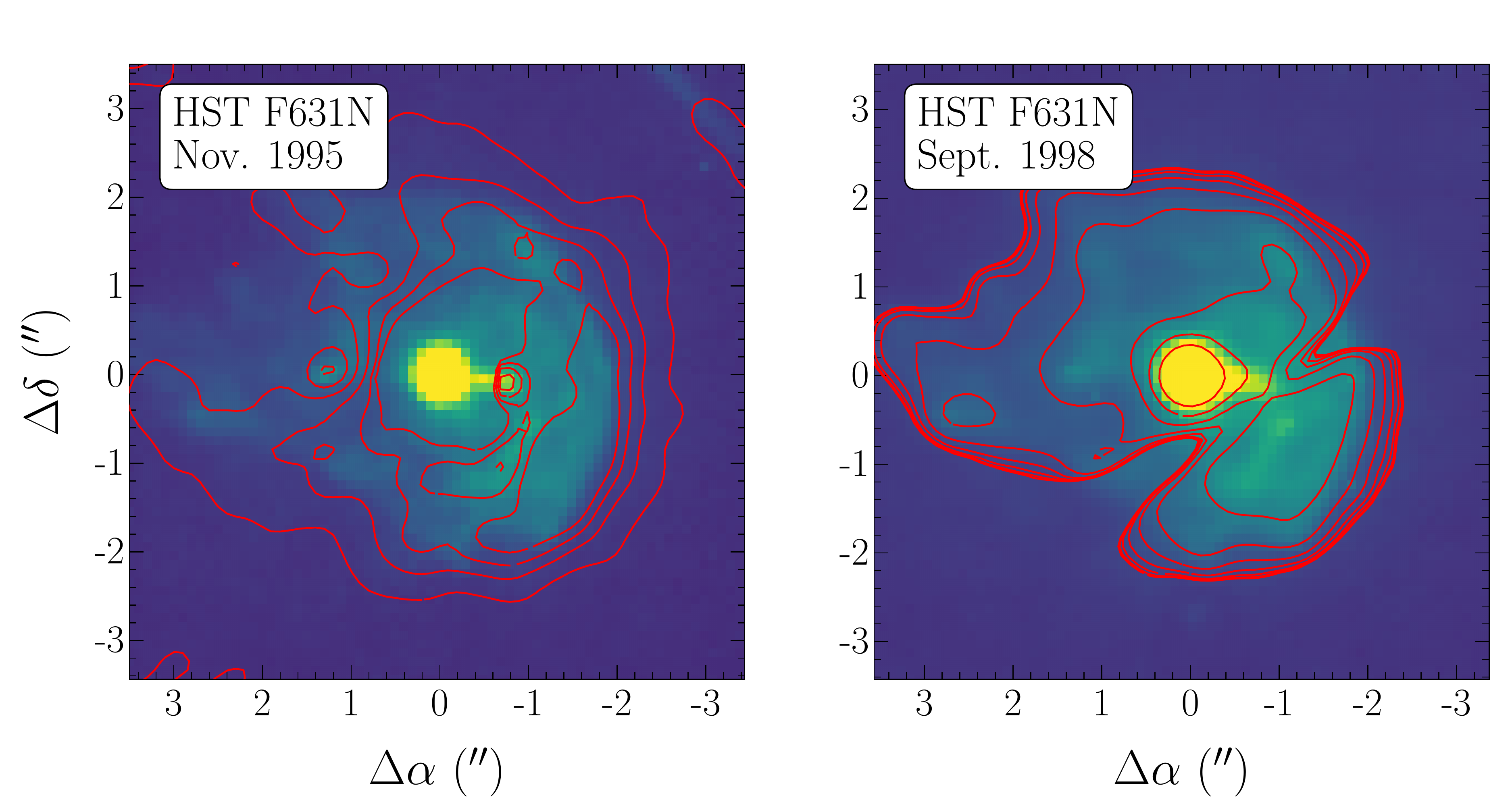}
   \caption{Median combined F631N images of the two HST epochs used in our proper motion study. The red contours are calculated from the residuals of the \Oi\, $-$ continuum images. The ``indented'' structure (right image) is due to the presence of a large diffraction spike in one of the stacked images.}%
   \label{fig:hst_epochs}
\end{figure}

We have identified six distinct knots with relatively strong S/N ($>10$), and clumpy emission signatures at a low S/N ($<5$) in at least two additional places. Not all knots are visible in our integrated channel maps due to integration over the entire emission range, while they are more clearly seen in smaller velocity channel maps. High velocity channels also show unresolved emission within 0\farcs3 ($\sim 120$~au) of the source. This is shown in Figures~\ref{fig:chanmaps01} and~\ref{fig:chanmaps02}. We have adopted a simple notation scheme, with each knot labeled according to its distance from the driving source, that is to say knot E1 is furthest from the source in the eastern direction while W3 is the knot closest to the source in the western direction. Eastern knots are redshifted and western knots are blueshifted (see Figure~\ref{fig:244_440_annotated}). We list the knot names and relative offsets from the source in Table~\ref{tab:244_440_knot_positions}.

\begin{table}
  \centering
  \caption{\label{tab:244_440_knot_positions} Knot positions}
  \begin{tabular}{crr}
  \hline
  \hline
  Knot & $\Delta\delta$~({\arcsec})& $\Delta\alpha$~({\arcsec}) \\ 
  \hline
  E1 & $-0.5018$ & 2.7272 \\
  E2 & $-0.5338$ & 2.2182 \\
  E3 & $0.0145$ & 1.4457 \\
  W3 & $-0.0793$ & $-0.6504$ \\
  W2 & $-$0.1006 & $-$0.8255 \\
  W1 & $-$0.1177 & $-$0.9649 \\
  \hline
  \hline
  \end{tabular}
  \tablefoot{Offsets of the primary knots relative to the source in 244-440 measured in the MUSE data. The source position is taken to be 0\arcsec. Knots labeled E$_n$ correspond to eastern, redshifted knots, while W$_n$ refers to western, blueshifted knots.}
\end{table}

\subsection{Comparison with HST observations for proper motion study}
\label{sub:hst_compare}

Our MUSE observations were compared with archival HST data to examine the time evolution of the outflow. Blue-shifted western jet emission is seen in the F631N and F673N filters, and redshifted eastern jet emission is faintly seen in the F631N filter. In both epochs of archival observations the western jet appears as a continuous stream, making proper motion estimations unreliable for those knots. However, the knots E1 and E3 are seen in the F631N images with the longer exposure times. 

\begin{figure}[htpb]
   \centering
   \includegraphics[width=\linewidth]{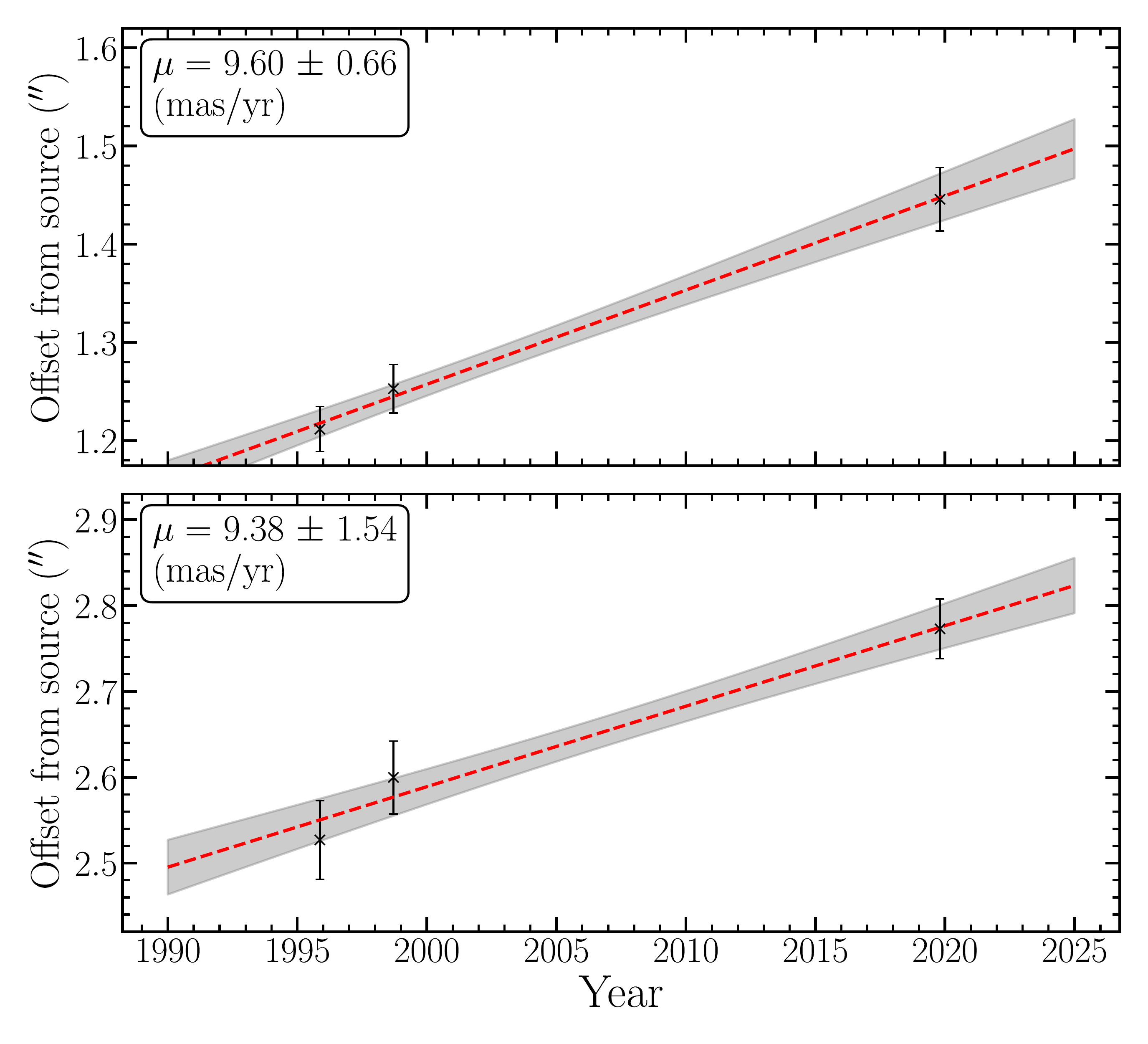}
   \caption{{Relative knot offsets for the three epochs of observations. A weighted least-squares fit is plotted in red and indicates a proper motion of $9.60 \pm 0.66$~mas yr$^{-1}$ for knot E3 (top) and $9.38 \pm 1.54$~mas yr$^{-1}$ for knot E1 (bottom). The filled gray bars indicate the 1$\sigma$ uncertainty of the fits.}}%
   \label{figure:pm_scatter}
\end{figure}

To estimate the positions of the redshifted knots, we performed centroid fitting to the knots E1 and E3 and to the stellar profiles in both the median \Oi\, and residual \Oi\, images using 2D Gaussian functions. The relative offsets in the x- and y-directions are listed in Table~\ref{table:offsets_and_pa}. There is the most uncertainty as to these offsets in the 1998 observations due to the low exposure times.
Additionally, knot E1 is very faint in both HST epochs, resulting in larger uncertainties also for its position. Figure~\ref{figure:pm_scatter} shows the relative knot offsets as a function of observation time and the best-fit line through the data, providing a proper motion of $ 9.5 \pm 1.1$ mas yr$^{-1}$, 
which corresponds to $\sim 15-19$~\kms. This is in agreement with typical tangential velocities of HH objects in the ONC~\citep[$< 50$~\kms; see][]{Reiter2016mnras} and consistent with the values normally found in low-mass stars and substellar objects~\citep{Whelan2014, riaz2017}. 

The radial velocity can be calculated from centroid fits of the knots at each emission line with respect to the ONC flow velocity, resulting in an average value of $v_\mathrm{rad} = 56 \pm 10$~\kms, and thus an absolute jet velocity of the same order, about 60~\kms.  The resulting jet inclination angle $i_\mathrm{inc} = \atan{v_\mathrm{rad} / v_\mathrm{tan}} =$ { \ang{72.2} $\pm$ \ang{4.2}} with respect to the plane of the sky is not consistent with the previous interpretations of a nearly edge-on disk, which would imply a much lower jet inclination. We discuss our interpretation of this in Section~\ref{sub:pm_implication}.

These values additionally allow us to estimate a lower limit on the dynamical age $t_{\rm dyn}$ of the knots. The knot E1 is the most distant knot with a deprojected distance of $\sim 2\farcs8$ ($\sim$1120 au\,) which yields $t_{\rm dyn} \sim 300$~yr.


\begin{table*}
\centering
\caption{\label{table:offsets_and_pa} Relative offsets and PAs}
\begin{tabular}{*{7}{c}}
   \hline\hline
   \multirow{2.5}{*}{Obs. Date}
      & \multicolumn{3}{c}{E1}
	 & \multicolumn{3}{c}{E3} \\
   \cmidrule(l){2-4} \cmidrule(l){5-7}
	 &  $\Delta\delta$ (\arcsec)& $\Delta\alpha$ (\arcsec) & P.A. (\degr) & $\Delta\delta$ (\arcsec) & $\Delta\alpha$ (\arcsec) & P.A. (\degr) \\
	 \hline
   1995 Nov 14 & $-$0.4358  & 2.4981  &  99.9  & 0.0497 & 1.2106   & 87.7 \\
   1998 Sep 17 & $-$0.4266  & 2.5646  &  99.4  & 0.0808 & 1.2502  & 86.3 \\
   2019 Oct 23 & $-$0.5018  & 2.7272  & 100.4  & 0.0145 & 1.4457  & 89.4 \\
   \hline\hline
\end{tabular}
\tablefoot{{Offsets} and PAs for knots E1 and E3 calculated from the three epochs of observations. {The offsets are measured relative to the source.}}
\end{table*}

\subsection{Jet curvature}
\label{sub:curvature}

Our data allow us to observe, for the first time, the redshifted jet and its strong curvature. The [\ion{O}{i}] line shows the morphology of the jet and the direction of the ionization front.
As $\theta^2$ Ori A is the dominant ionization source east of the Orion Bar~\citep{ODellApJ2017}, one would expect it to contribute the greatest radiation. 
The radiation contribution from $\theta^2$ Ori B is not as powerful but is likely not negligible. Indeed, in Figure~\ref{fig:244_440_annotated} we observe knot E3 in the redshifted jet with what may be a bow shock pointing approximately \ang{50} north through east, congruent with a wind from Ori B, and the curvature traced by knots E2 and E1 appears to be in agreement with a strong wind from $\theta^2$ Ori A that is deflected in part by a lesser wind from $\theta^2$ Ori B. 

Additionally, we see in Figure~\ref{fig:244_440_annotated} a distinct asymmetry in outflow scale, with the redshifted jet having a higher displacement than its blueshifted counterpart. As the eastern jet flows downstream with respect to $\theta^2$ Ori A, it may be that crossing the photoionization front and associated shocks, or even puncturing through the envelope, highly disturbs the western jet. 

\section{Physical properties of the jets}
\label{sec:physical_properties}

The environment of the Orion region presents a unique opportunity to examine the physical conditions of outflows and compare what we can observe in irradiated proplyds with what is known about more isolated outflow conditions. The electron densities and electron temperatures in the proplyd structure are key elements for determining the mass-loss rates of proplyds, and are integral in understanding the nature of stellar evolution~\citep{RayNewAR, reipurthbally2001}. Accurate flux measurements are thus necessary to explore the physical conditions of the proplyd jets. In this section we discuss the role of extinction and examine emission line ratios to explore what diagnostic tools we can apply to this object.

\begin{figure}[t]
  \centering
  \includegraphics[width=\linewidth]{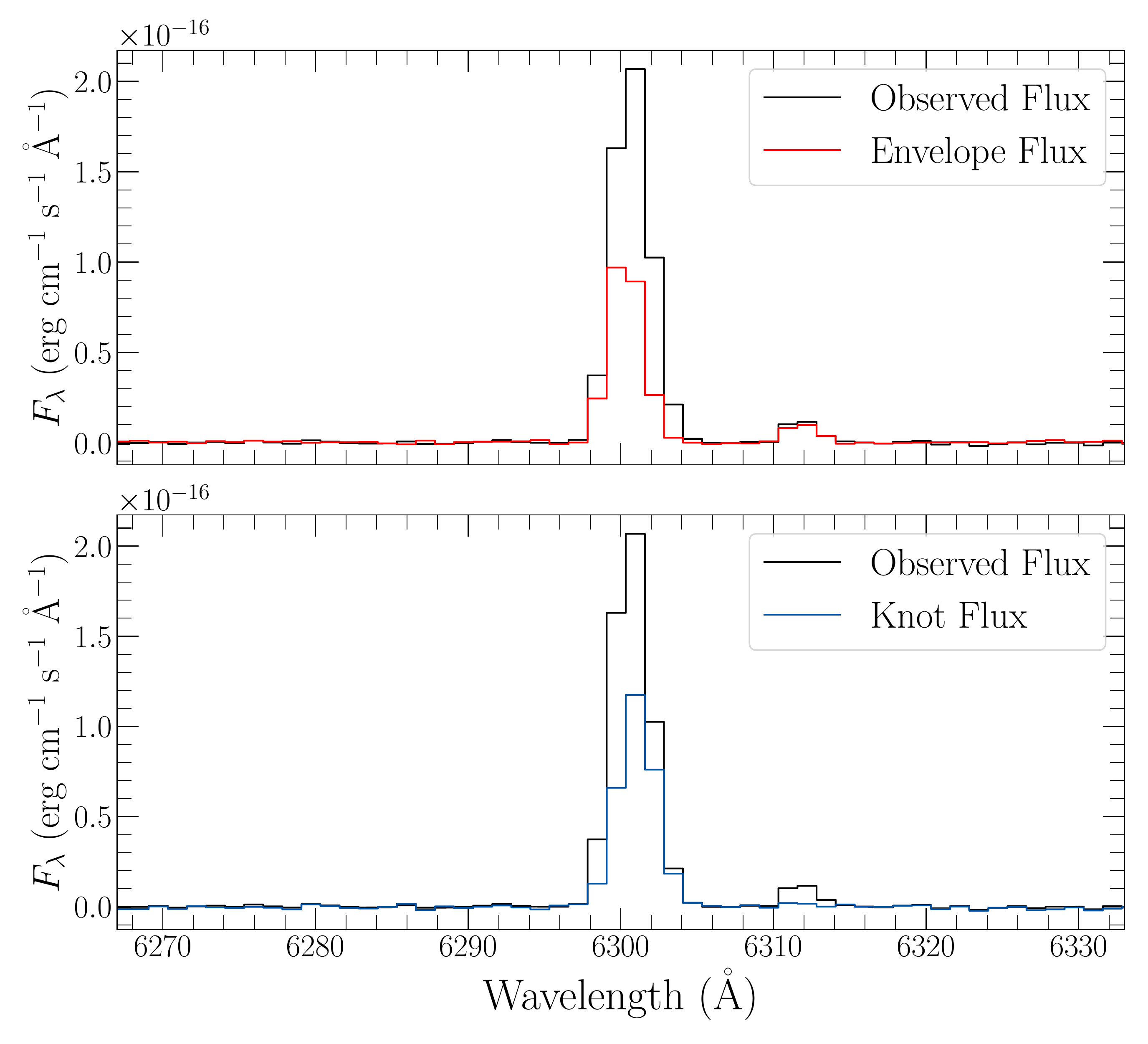}
  \caption{Spectral profiles of the observed flux, envelope flux, and intrinsic knot flux. The top panel compares the observed and envelope contributions, while the bottom shows the observed and intrinsic residual. The spectra are of knot E3 extracted from the \Oi\, line.}%
  \label{fig:oi_obs_env_knot}
\end{figure}

In this analysis, we extracted knot fluxes using circular apertures ($r = 3$~pixels or 0\farcs075) centered at the positions given in Table~\ref{tab:244_440_knot_positions}. These fluxes were then corrected for extinction, as discussed in Section~\ref{sub:extinction}. In lines where the observed flux is a combination of the knot and envelope emission, we used apertures of the same size to extract fluxes nearby in the envelope to estimate the proplyd contribution and subtract these from the knots. An example of this is shown in Figure~\ref{fig:oi_obs_env_knot}.

As discussed above, the physical conditions in the jet are key to understanding the evolution of the star. The difficulty posed by proplyds, however, is that their densities often exceed the critical limits of traditional diagnostic ratios, such as [\ion{S}{ii}], [\ion{O}{i}], and [\ion{N}{ii}]~\citep[see][for a deeper discussion]{Bally1998AJ, Henney1998AJ, MesaDelgado2012MNRAS}. Additionally, such techniques as those proposed by~\citet[][the so-called BE99 method]{bacciottieisloffel}  rely on ratios relative to \Ha. Because of the strength of the envelope emission in \Ha\, in our data, it was not possible to recover the knot fluxes in this line, and so we could not rely on the BE99 technique.

Alternative density-sensitive lines are seen in our data, primarily forbidden Fe lines, which can be a powerful diagnostic tool for tracing denser emission regions~\citep{podio2006}. Additionally we observe [\ion{Ni}{ii}] emissions, which can be used to probe fluorescent excitation as well~\citep{Lucy1995A&A, GianniniApJ2015b}.

\subsection{Extinction}%
\label{sub:extinction}

Extinction plays a critical role in the analysis of the proplyd. Its quantification is further complicated by the role of extinction in the envelope of the proplyd as well as the photoevaporated and photoionized flow itself~\citep{Henney1999AJ, MesaDelgado2012MNRAS}, and the difficulty in determining what amount of nebular emission occurs in the forefront of or behind the proplyd envelope. Since the dust within the proplyd is not well known, these issues make disentangling the intrinsic propyld emission from the nebular cloud --- and by extension, the jet from the envelope --- a difficult procedure.

\begin{figure}[htpb]
  \centering
  \includegraphics[width=\linewidth, trim={0cm 1cm 0cm 0cm}, clip=true]{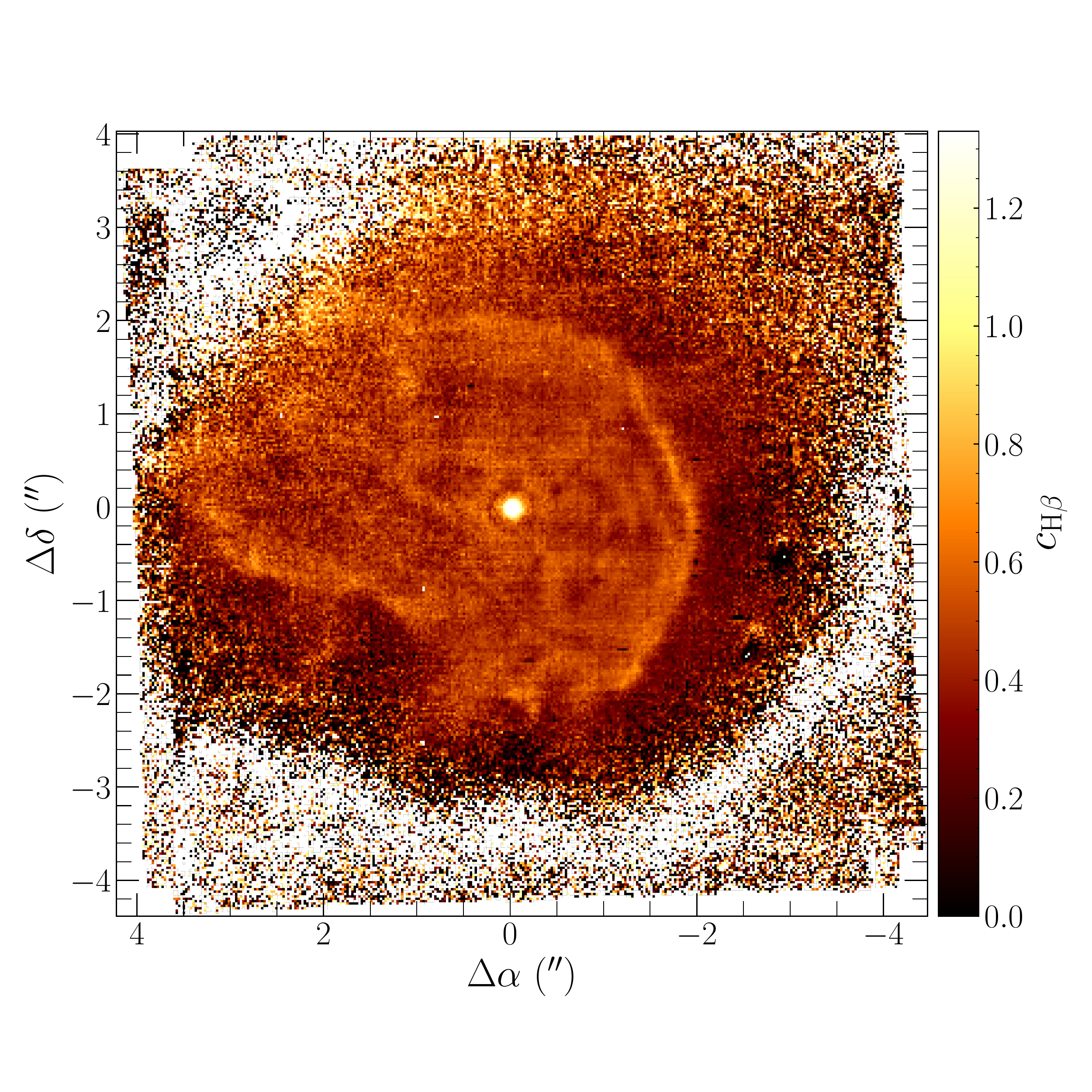}
  \caption{Spatial distribution map of the line-of-sight \chb\, coefficient for proplyd 244-440.}
  \label{fig:extinction_map}
\end{figure}

A few methods exist to explore this phenomenon, but for more general purposes we examined the simplest case. Using integrated channel maps of the Balmer lines (\Ha\, and \Hb\,), we constructed maps of the extincted dust in the FOV. This ratio was chosen due to its weak density-dependence, making it well-suited to the analysis of the dense inner envelopes of the proplyds~\citep{MesaDelgado2012MNRAS}. We utilized the reddening curve of~\citet{Cardelli1989ApJ} adapted for the optical regime by~\citet{ODonnell1994ApJ} and assumed $R_V=5.5$~\citep{Weilbacher2015AAm, McLeod2016MNRAS}. We further assumed under case-B assumptions an intrinsic \Ha$/$\Hb\, flux value of 2.86~\citep{McLeod2015MNRAS} and used the observed \Ha$/$\Hb\, ratio to produce pixel-by-pixel mappings of the \chb\, reddening coefficient along the line of sight in \texttt{PyNeb}\footnote{http://research.iac.es/proyecto/PyNeb/} (see Figure~\ref{fig:extinction_map}).

What we observe is that there is a clear morphological difference between background extinction and extinction due to dust in the proplyd envelope. In the case of 244-440, the envelope displays a clumpy but still fairly uniform structure, with the IF having the highest extinction. Examining the extinction in adjacent regions and within the proplyd we generally observe a higher extinction in the proplyd than in the nebula, which is in line with anticipations for a dusty envelope.  

With this map, we selected regions adjacent to and within the proplyd to calculate a representative \chb\, value. Under the assumption that foreground and adjacent background extinction is uniform, we estimated the intrinsic proplyd extinction $c(\mathrm{H}\beta)_p$ as the difference between the nebular extinction (\chb=$0.44-0.45$) and that of the proplyd envelope (\chb$\simeq0.6$) such that $c(\mathrm{H}\beta)_p = 0.15$. This intrinsic reddening coefficient is similar to what is found by~\citet{MesaDelgado2012MNRAS} for other proplyds closer to the Trapezium.  

Due to this small intrinsic extinction, we simplified our correction by taking a constant \chb\, value of 0.45 for our calculations. To eliminate the nebular contribution, we selected a sample of background regions away from the proplyd envelope and constructed a mean representative spectrum, which was subtracted from every pixel in the continuum-subtracted cubes. 

\begin{figure}[htpb]
  \centering
  \includegraphics[width=\linewidth]{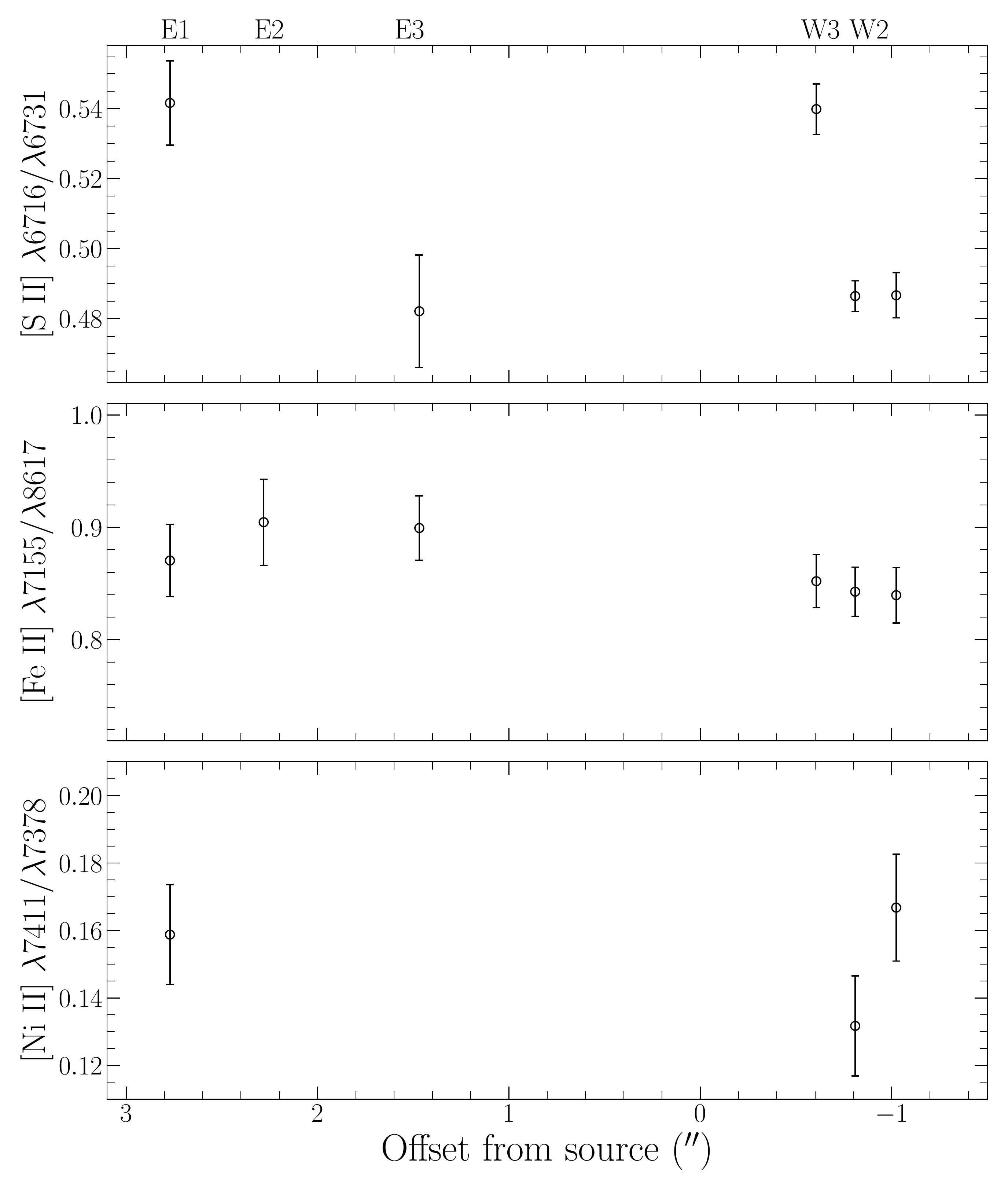}
  \caption{Flux ratios of the [\ion{S}{ii}] (top), [\ion{Fe}{ii}] (center) and [\ion{Ni}{ii}] (bottom) lines as a function of distance from the source. The zero-point represents the driving source. The fluxes are summed over circular apertures with a 3-pixel radius.}%
  \label{fig:ratio_vs_distance}
\end{figure}

\begin{figure}[htpb]
   \centering
  \includegraphics[width=\linewidth]{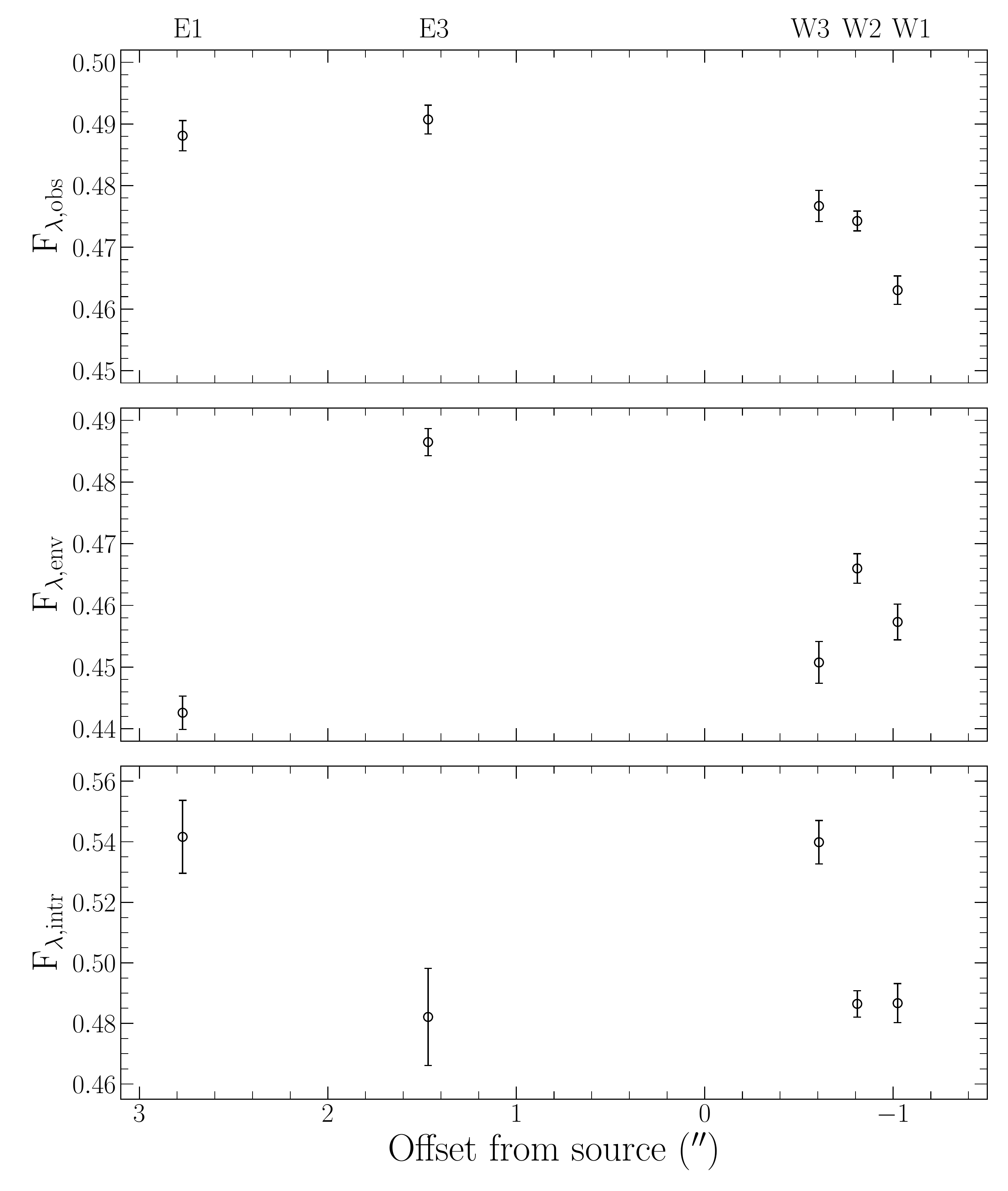}
   \caption{Ratios of the [\ion{S}{ii}] emission line for the total observed flux (top), envelope flux (center), and jet flux (bottom). The jet flux is estimated by subtracting nearby envelope flux from the total observed flux and correcting for the intrinsic proplyd extinction.}%
   \label{fig:sii_flux_ratios}
\end{figure}

\subsection{Shock versus photoionization}%
\label{sub:physical_conditions}

\begin{figure}[htpb]
   \centering
    \includegraphics[width=\linewidth]{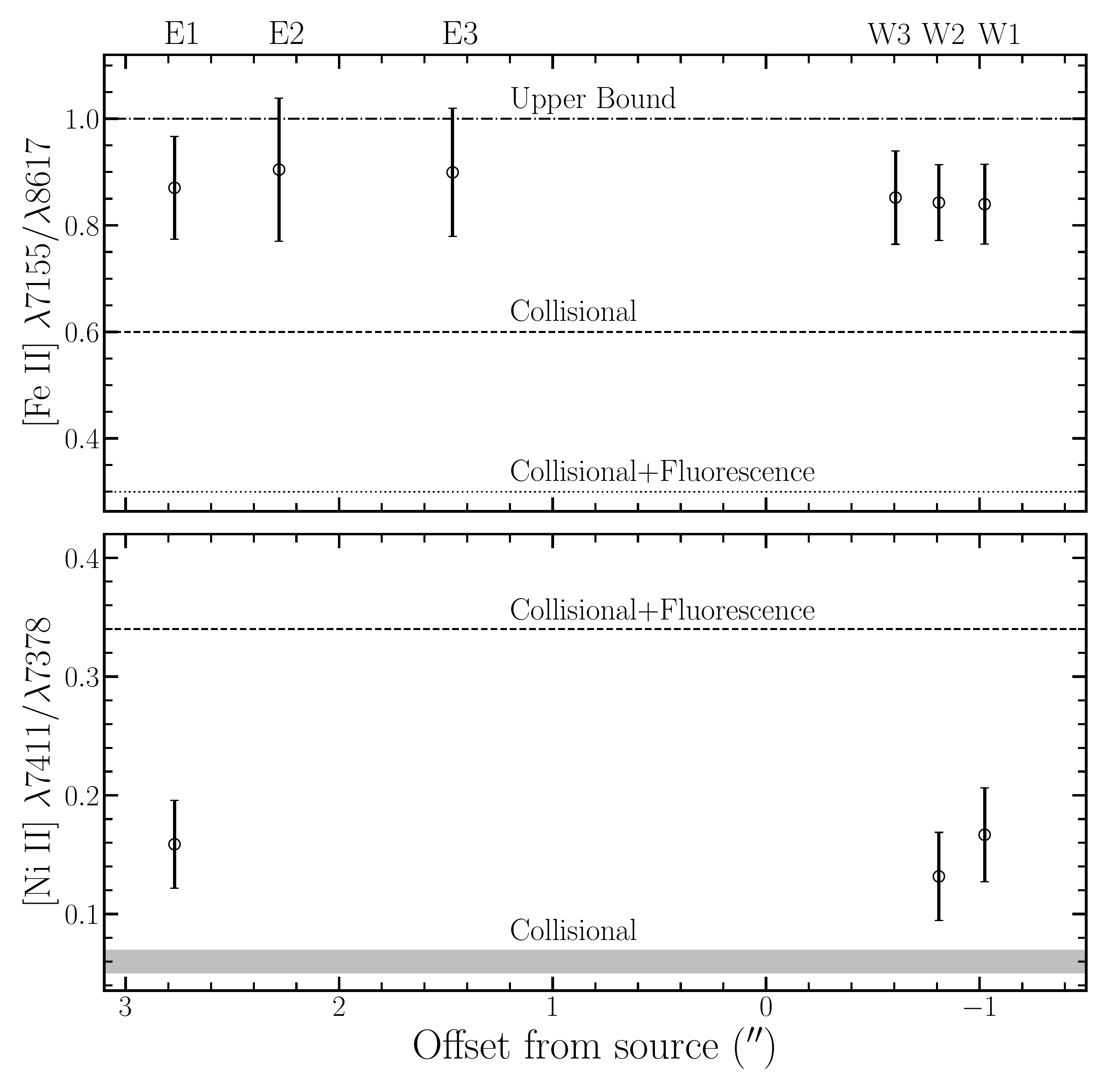}
   \caption{Emission line ratios of the [\ion{Fe}{ii}] and [\ion{Ni}{ii}] lines sensitive to fluorescent pumping. Horizontal lines indicate the predicted ratios for each case, taken from~\citet{GianniniApJ2015b}.}%
   \label{figure:fluor_plot}
\end{figure}

 We used the fluxes measured in Table~\ref{tab:244_440_fluxes} to estimate line ratios and compared them with those given in Table 2 of~\citet{GianniniApJ2015b} to test for fluorescent excitation. The predicted collisional$+$fluorescent excitation ratio for the [\ion{Ni}{ii}]$\lambda7411/7378$ line is given as 0.34, while the collisional case is given in the range 0.05$-$0.07; in our data we see ratios from 0.13$-$0.17. The [\ion{Fe}{ii}]$\lambda7155/8617$ ratio is $0.8-0.9$, which is higher than both collisional and collisional$+$fluorescence predictions.\footnote{The $\lambda7453$ line is omitted here due to the low S/N.} These are shown in Figure~\ref{figure:fluor_plot}. These results indicate that while fluorescent pumping does appear to enhance the emission it is not the dominant process, which is expected as optical lines like [\ion{Fe}{II}] are more likely to arise from collisional excitation than photoexcitation~\citep{BautistaApJ1996a}.

\begin{figure}[htpb]
   \centering
  \includegraphics[width=\linewidth]{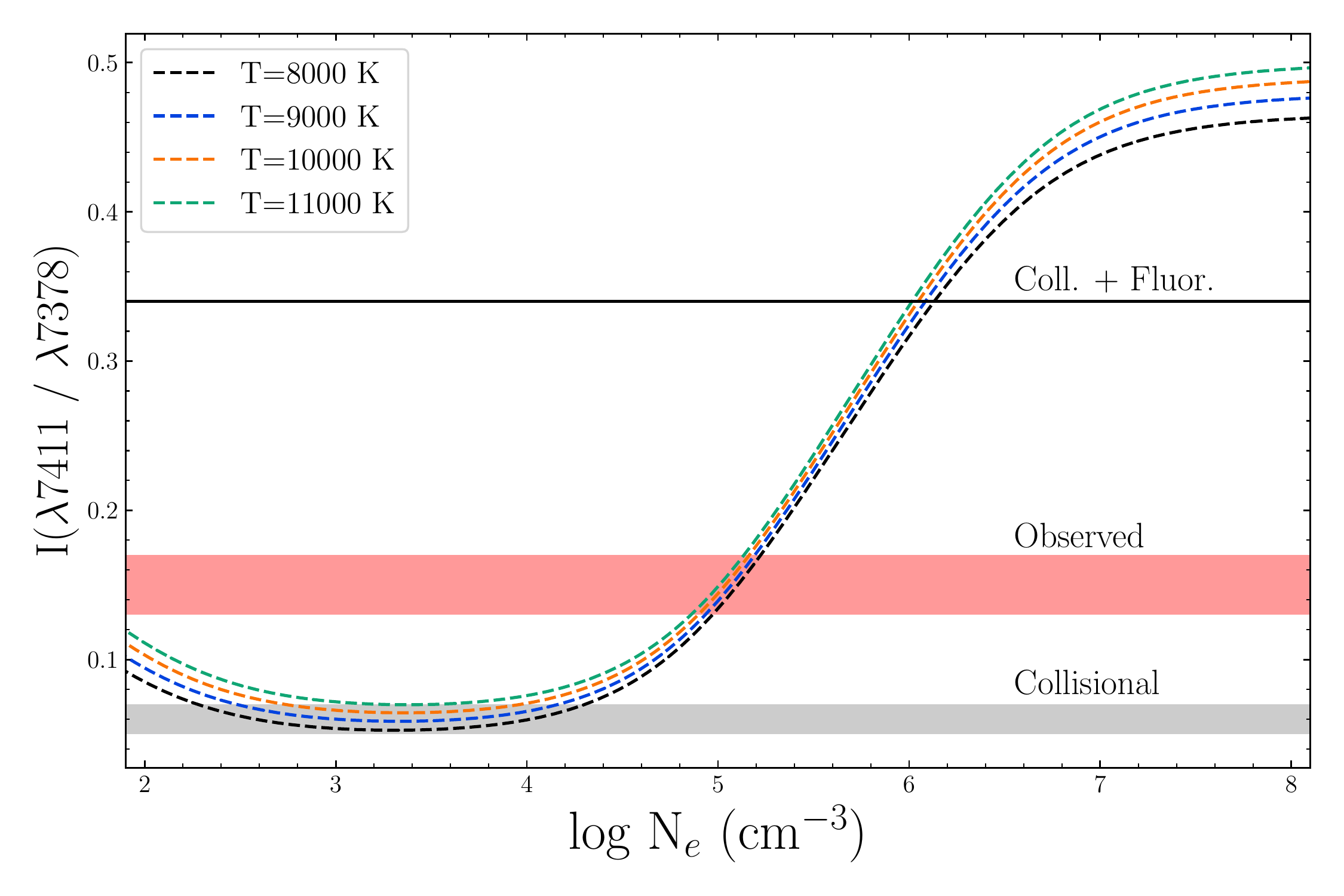}
   \caption{Emissivity ratio of the optically forbidden Ni emission line, calculated in \texttt{PyNeb}. The dashed curves are the emissivity ratios calculated for each given temperature. The gray region denotes the predicted range of ratios for collisional excitation. The red region denotes the range of values we observe in our data. The solid black line denotes the predicted ratio for collisional$+$fluorescent excitation. The predicted values are taken from~\citet{GianniniApJ2015b} assuming $T_e = 10,000$~K and $n_e \sim 10^{3}-10^{5}$~cm$^{-3}$.}%
   \label{fig:pyneb_ni2_ratio}
\end{figure}

\subsection{Mass-loss rate in the jet}%
\label{sub:mass_loss}

In Section~\ref{sub:physical_conditions} we discussed the difficulty posed by the general densities of proplyd envelopes. Similarly, the lack of jet emission in temperature-sensitive lines further complicates traditional diagnostic techniques, introducing uncertainties to our determination of the electron density. However, we note that the mass-loss rate is only weakly dependent on the electron density and so we set reasonable bounds on this physical parameter. We make estimates based on Figure~\ref{fig:pyneb_ni2_ratio}, where our observed [\ion{Ni}{ii}] ratios imply electron densities $7\times10^4$ cm$^{-3} \leq n_e \leq 2\times10^5$ cm$^{-3}$. Our [\ion{Fe}{ii}] ratios similarly {imply electron densities $\sim$ 10$^5$ cm$^{-3}$.} As the [\ion{Fe}{ii}] and [\ion{Ni}{ii}] ratios for the observed knots exhibit relatively small scatter in their respective emission lines (see Figure~\ref{figure:fluor_plot}), we assume a mean density of $n_e = 1.3 \times 10^5$~cm$^{-3}$ at $T_e = 8200$~K. 

Following~\citet{HartiganApJ1995}, we used the intrinsic knot luminosity of the \Oi\, and [\ion{S}{ii}]$\lambda6731$~\AA\, lines to estimate the jet mass and outflow rate.  
The mass outflow (in units of \Msun\, yr$^{-1}$) is given by the equation

\begin{equation}
   \centering
   \label{eq:oi_mass_loss}
   \begin{split}
      \dot{M}_{out} &= 2.27 \times 10^{-10} \left( 1 + \frac{n_c}{n_e} \right) \left( \frac{L_{6300}}{10^{-4}~L_{\odot}} \right) \\
		    & \times \left( \frac{v_\mathrm{tan}}{150~\mathrm{km~s}^{-1}} \right) \left( \frac{\ell_\mathrm{tan}}{2 \times 10^{15}~\mathrm{cm}} \right)^{-1} M_{\odot}~\mathrm{yr}^{-1}
   \end{split}
\end{equation}
 
\noindent for the \Oi\, line, where $n_c$ is the critical density, $n_e$ the electron density, $v_\mathrm{tan}$ the proper motion (in \kms), and $\ell_\mathrm{tan}$ is the size of the aperture (in cm) in the plane of the sky. The expression is similar for the [\ion{S}{ii}]$\lambda6731$ line,

\begin{equation}
   \centering
   \label{eq:sii_mass_loss}
   \begin{split}
      \dot{M}_{out} &= 3.38 \times 10^{-8} \left( \frac{L_{6731}}{10^{-4}~L_{\odot}} \right) \left( \frac{v_\mathrm{tan}}{150 \mathrm{km~s}^{-1}} \right)\\
		    & \times  \left( \frac{\ell_\mathrm{tan}}{2 \times 10^{15}~\mathrm{cm}} \right)^{-1} M_{\odot}~\mathrm{yr}^{-1}
   \end{split}
\end{equation}

\noindent where it is assumed to be in the high-density limit so that the ratio $n_c/n_e \ll 1$. 

For these calculations, we used a tangential velocity $v_\mathrm{tan} = 17$~\kms\, and an aperture size~$\ell_\mathrm{tan} = \SI{9.0e14}{\centi\meter}$. We find low luminosities and mass-loss rates for all of the knots, with the blueshifted jet presenting higher values due to its exposure to the ionizing winds. For the \Oi\, lines the mass-loss rate is on the order of $10^{-11}$~\Msun\, yr$^{-1}$, while in the [\ion{S}{ii}] line it is on the order of $10^{-10}$~\Msun\, yr$^{-1}$~(see Table~\ref{tab:luminosities}). 

We note however that these values do not account for photoionization effects, and so represent a lower limit of {the jet} mass outflow rate. If instead we assume photoionization plays a large role in the jets, then we can calculate an upper limit of the mass-loss rate in the jet from

\begin{equation}
   \centering
   \label{eq:bally_mass_loss}
   \begin{split}
      \dot{M}_j = 3.4 \times 10^{-9} \left( \frac{v_j}{100~\mathrm{km~s^{-1}}} \right) \left( \frac{n_e}{10^{3}~\mathrm{cm}^{-3}} \right) \\
      \times \left( \frac{r_j}{115~\mathrm{au}} \right)^3 M_{\odot}~\mathrm{yr}^{-1}
   \end{split}
\end{equation}

\noindent as given in~\citet{Bally2001ApJ}. With this equation, we estimate that $\dot{M}_j$ would have an upper limit of $10^{-9}$~\Msun\, yr$^{-1}$. 


\begin{table}
  \centering
  \caption{\label{tab:luminosities}Jet luminosity, mass, and outflow rates}
    \begin{tabular}{
    c
    *{6}{S[table-format=2.2]}
    }
      \hline\hline
      \rule{0pt}{3ex}
      \multirow{2.5}{*}{Knot}
	 & \multicolumn{3}{c}{[\ion{O}{i}]$\lambda6300$}
	    & \multicolumn{3}{c}{[\ion{S}{ii}]$\lambda6731$} \\
      \cmidrule(l){2-4} \cmidrule(l){5-7}
	    & $L$  & $M$ & \ensuremath{\dot{M}_{out}} & $L$  & $M$ & \ensuremath{\dot{M}_{out}} 
	    \rule{0pt}{-3ex}\\
	    \hline
	    \rule{0pt}{3ex}
      E1	& -5.2 & -9.0 & -10.5 &      &        &      \\
      E2	& -5.6 & -9.5 & -10.9 &      &        &      \\
      E3	& -5.2 & -9.0 & -10.5 & -5.7 &  -9.7  & -9.9 \\
      W3	& -4.8 & -8.6 & -10.1 & -5.0 &  -9.0  & -9.3 \\
      W2	& -4.7 & -8.6 & -10.0 & -4.9 &  -8.9  & -9.2 \\
      W1	& -5.0 & -8.9 & -10.3 & -5.0 &  -9.0  & -9.3 \\
      \hline\hline
   \end{tabular}
   \tablefoot{Luminosity, mass, and outflow rates for the redshifted knots in the \Oi\, and [\ion{S}{ii}]$\lambda6731$ emission lines. Luminosity is given in units of {$\log_{10}$}~$L_{\odot}$, masses in units of {$\log_{10}$}~\Msun, and mass-loss rates in units of {$\log_{10}$}~\Msun~yr$^{-1}$.}
\end{table}

\section{Discussion}
\label{sec:discussion}

\subsection{Implications of the proper motions}
\label{sub:pm_implication}
\subsubsection{Constraints on disk inclination}
\label{subsub:inconsistencies}

Throughout our study we rely largely upon proper motion measurements to constrain the jet velocity. This parameter is important for both considerations of the wiggling jet model, and for calculations of the jet mass-loss rate. In Section~\ref{sub:hst_compare} we found proper motions $<20$~\kms, and estimated a jet inclination angle $i_\mathrm{inc}$ of {\ang{72.2} $\pm$ \ang{4.2}}. While the low proper motion is not unusual for objects in the ONC, we note that $i_\mathrm{inc}$ is inconsistent with the observations of~\citetalias{Bally2000}, which the authors interpret as a nearly edge-on disk. We examine a few possibilities for this below.

Our assumption in the text is that the jet moves through the medium in a ballistic manner, that is, with a constant velocity. If the jet encounters a dense material, it may be deflected and result in a change in its bulk flow velocity. A similar phenomenon was recorded by~\citet{hartiganholcomb2019} in the objects HH 8 and HH 10, where stationary ``loop'' structures occurred in the knots as they interacted with sheets of ambient material, resulting in no measurable proper motions. If there are dense, unseen obstacles within the proplyd envelope along the flow path of the jet, or if the jet is piercing through the envelope, it may be possible that the jet is significantly slowed or deflected as it encounters the material, or that the unshocked or weakly shocked material in the jet produces shocks in its vicinity that falsely present as knots. The latter case would raise the possibility that the locations of the knots observed in the HST data are not knots at all, but density enhancements in the surroundings. In such a scenario our proper motions would not be reliable as a way to estimate the inclination angle of the jet, and we would require measurements closer to the driving source that are more likely to be unaffected by the above interactions. However, we note in Figure~\ref{fig:extinction_map} that while there is obvious structure in the density distribution of the envelope, it does not appear sufficient to deflect the jet in this way. It is important to note nonetheless that the blueshifted jet's proximity to the IF may affect a change in the jet structure as it passes through the front. If this is the case, the ballistic assumption may not be appropriate for the blueshifted jet. 

In the HST images, particularly the \Ha\, filter, \citetalias{Bally2000} observed a silhouetted structure which they interpreted as an edge-on disk (see their Figure 7b) with a semi-major axis nearly aligned vertically in the image. In our MUSE observations we observe a bright ``halo'' structure around the source in~\Ha\, and~\Oi\, but we are unable to determine its cause. If we look solely at the properties of the jet, we conclude that it does not support an edge-on disk scenario. Moreover, if the jet is close to the plane of the sky, the radial velocities would imply a driving source larger than the one supported by its spectral type (see Appendix~\ref{sec:appendixA}). Additionally, the large gap between the source and the nearest knot E3 in the redshifted flow versus the smaller gap observed in the blueshifted side suggests a system where the size and orientation of the disk effectively obscures the receding jet. Finally, the jet direction is not centered with the disk axis.

We argue that all these discrepancies are reconcilable if one assumes that the disk observed in the HST images is not associated with the jet-driving source. Such a scenario would explain the position angle of the jet, its low proper motion, and our derived jet inclination angle. In our discussion on the jet curvature below, we see that this conclusion is also a possible outcome of our modeling and consistent with previous observations.

\subsubsection{Dynamical age of the jet}
\label{subsub:jet_age}

In Section~\ref{sub:hst_compare} we used the best-fit proper motion of the jet and the offset of knot E1 from the source to present a minimum age of the jet of at least 300~yr. This is a most interesting result, as it tells us not only that the jet is quite young and still active, but also that photoionization and photoevaporation of the proplyd do not seem to affect the ability of the star to launch a jet. The jet may have been launched within a ``bubble'' that shielded it from the majority of hard-UV radiation, which may explain why collisional excitation appears to dominate in the jet.

Additionally, we note above (Sec.~\ref{subsub:inconsistencies}) that the proximity of the blueshifted jet to the IF may change the jet structure. Along with the age estimate, this raises questions about the length of the blueshifted emission, which is substantially shorter than that of its redshifted counterpart. The measured radial velocities of the knots are not too dissimilar between the red and blue lobes, suggesting similar launch velocities. If we further assume similar launch epochs, then we would expect the blue lobe to extend to at least the same length, yet we observe no emission beyond the envelope of the proplyd. Furthermore, YSOs typically begin driving outflows early in their life-cycles~\citep[as early as $10^4$~yr; see][]{andre2000PPIV}, so it is expected that the true extent of the jet may be far greater than what is observed~\citep{frank2014}. This is not unreasonable, and we raise two points here to address this.

Firstly, visible knots are formed by shocks as the jet interacts with itself or the ambient medium (in isolated regions), or by illumination from an external source (in irradiated regions). The minimum age of the star suggests that the jet is far older than 300~yr, so the jet must have punctured through the envelope and passed into the larger region of the nebula, which in the neighborhood of our object of interest typically has a density on the order of $10^3$~cm$^{-3}$~\citep{MendezDelgado2021}. Other HH objects are seen in the ONC beyond the protective shells of the proplyds, which indicates that the ONC is either dense enough or irradiated enough to render these objects visible. It may be that the jet has lost so much of its density after passing beyond the envelope that it is simply not visible, which is reasonable as the luminosities of HH objects tend to decrease with separation from the driving source. Secondly, the environment surrounding a proplyd is not hospitable to transient objects. Given the low luminosity and radial velocities of the jets we do not anticipate a very strong driving force, so it is likely that the strong stellar winds and radiation in the ONC have entirely dissipated the portions of the jet beyond the envelope. 

\subsection{Origin of the curvature}
\label{sub:origin_curvature}
To explain the C-shaped symmetry common to irradiated jets in the Orion nebula~\citep[][and references therein]{Bally2001ApJ}, a few models have been put forth involving ram pressure from stellar winds~\citep{Raga2009AA, Estalella2012AJ}. Additionally, sinusoidal jet morphologies may arise due to the presence of a binary companion which causes either orbital motion of the jet source~\citep{masciadri2002, Lai, Murphy2021, Erkal2021AAa}{,} or a precession of the jet ejection axis due to the inner disk not being coplanar with a companion's orbit~\citep{ZhuMNRAS2019, terquem1999}. \citetalias{Bally2000} initially proposed the idea of a hidden companion based on the offset of the photometric center from the geometric center of the disk as seen in their [\ion{O}{i}] observations. Recent evidence put forth by~\citet{Tobin2009ApJ} and~\citet{Kounkel2019AJ} also indicates that 244-440 is a spectroscopic binary. The possibility of 244-440 possessing a companion that may produce a ``wiggling'' in the jet axis, as well as the ram pressure from the stellar winds from the stars in the Trapezium cluster and $\theta^2$ Ori A and B, may all act in conjunction to produce the complex morphology that we observe in this object.

In this section we explore whether the observed curvature in the jet can be explained by a wiggling jet model in the absence of an appreciable side-wind. The basic parameters of these models are the length scale of the wiggle $\lambda$ and the half-opening angle of the jet cone $\beta$, {an important parameter in the precession model}. They can be inferred by visual inspection of the data and the proper motion estimates presented in Section~\ref{sub:hst_compare}. We measured the knot centroid positions by Gaussian fitting of the jet along the outflow axis, and utilized the jet inclination angle estimated above to de-project these positions from the plane of the sky. In the following analysis, we follow the method explored by~\citet{Murphy2021}.

From our data, we estimate a de-projected length-scale $\lambda \sim $8\arc. The half-opening angle $\beta$ is also observable from the data by fitting the slope of the peaks of the wiggle curve. Even if we do not observe as many peaks as some other wiggling jets, we can safely estimate $\beta= $\ang{3.4}. We can further relate the precession model to the orbital model by means of $\beta$. In an orbital model, we define the ratio of the orbital velocity to the jet velocity as $\kappa = v_o / v_j$, and this is related to $\beta$ by $\kappa \leq \tan \beta$. 

\begin{figure}[htpb]
  \centering
  \includegraphics[width=\linewidth]{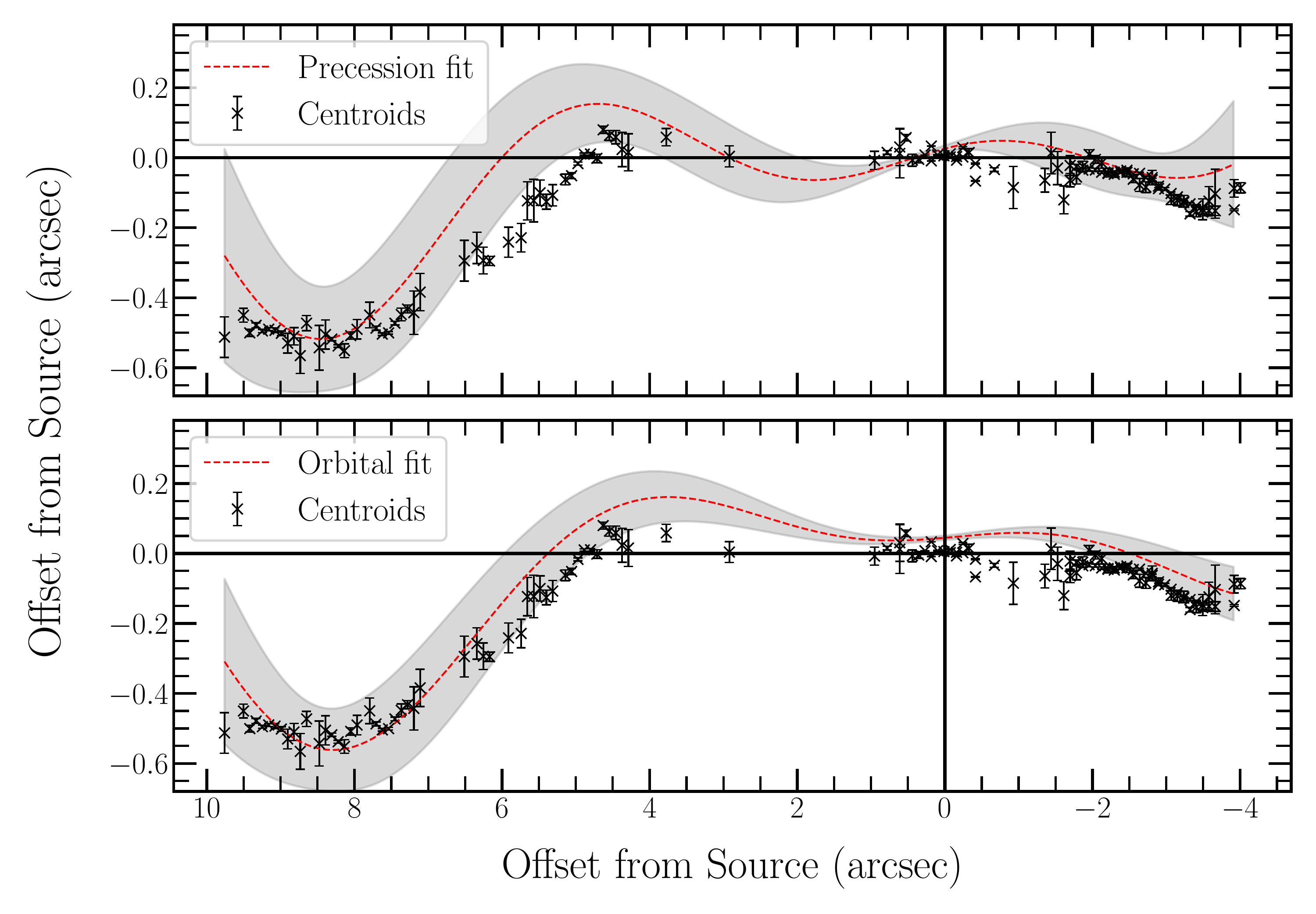}
  \caption{Best fits for the precession (top) and orbital (bottom) models for the [\ion{Fe}{ii}]$\lambda7155$~\AA\, jet. The relative offset along the jet axis is deprojected by $x^{\prime} = x/\cos\theta$ for an angle of \ang{73} assuming a distance $d=400$~pc. The gray region indicates a 3$\sigma$ uncertainty in the model.}%
  \label{fig:prec_orb_fit}
\end{figure}

Using these parameters and the equations shown in~\citet{Murphy2021} we explored both an orbital motion model and a precession model for the jet. We used the Python package \texttt{lmfit}\footnote{https://lmfit.github.io/lmfit-py/} and rewrote the equations as functions of $\lambda$ and $\beta$, allowing them to vary by about 15\%. We used a fixed value of $v_j = 65$~\kms\, for the jet velocity and set an upper limit on $r_o$ of 0\farcs1 (see Section~\ref{sec:intro}). The results of the fits are shown in Figure~\ref{fig:prec_orb_fit}. The errors on the centroids were calculated according to Equation A.1 in~\citet{Porter2004}. Both models were weighted with the centroid errors.

It is important to recognize that given the short length-scale of the blueshifted jet and the inability to estimate a proper motion for that emission lobe, it is difficult to determine whether the wiggle is point-symmetric (precession) or mirror-symmetric (orbital motion) around the origin, though Figure~\ref{fig:prec_orb_fit} does favor mirror-symmetry. Nevertheless we can make assumptions based on the values derived from these fitted models.

A primary driver of the values derived from the fitted curves is the total system mass. The spectral type of the visible central star implies a low-mass object (see Appendix~\ref{sec:appendixA}) and we therefore limit our considerations to \Msys~$\leq 0.8$ \Msun. For the orbital motion model, the results imply $\mu$ ($= M_\mathrm{c}/M_\mathrm{sys}$) values ranging from $0.7$ with \Msys$=0.8$~\Msun\, to $0.9$ with \Msys$=0.4$~\Msun. We can estimate the maximum binary separation $a$ as the ratio of the orbital radius $r_o$ of the source about the barycenter and the mass ratio $\mu$,  $a = r_o/\mu$, obtaining separation in the range $\sim 30-40$~au. The implied orbital period $\tau_o$ is found by

\begin{equation}
    \centering
    \frac{M_\mathrm{sys}}{M_\odot} = \mu^{-3} \left( \frac{r_o}{\mathrm{AU}}\right)^3 \left( \frac{\tau_o}{\mathrm{yr}} \right)^{-2}
\end{equation}

\noindent which suggest an orbital period of $\sim 220$~yr. 

A similar calculation for the precession model results in a precession period $\tau_p = 238$~yr with $\mu$ ranging from $0.3$ (\Msys$=0.8$~\Msun) to $0.5$ (\Msys$=0.4$~\Msun)  
These results imply orbital periods of $\sim 30-50$~yr and binary separations from $\sim5-10$~au. This smaller value is not unreasonable as the precession model requires a warped inner disk, and if this is induced by a companion then the companion must be relatively close. 

In the precession model, the inferred mass-ratio indicates an equal-mass binary at close separation. The length scale of the deprojected jet ($\sim 4000$~au) and large fluctuation in the red lobe are consistent with precession models, as fluctuations due to orbital motion are more likely to appear within 100s of au of the source~\citep{Masciadri2001AJ}. However, this model presents some difficulties. If the companion is close in mass to the primary, one should wonder whether or not this companion would truly be hidden in observations. Unfortunately, we cannot resolve the binary separations in our MUSE observations, nor is our spectral resolution high enough to perform accurate spectro-astrometry. If we consider larger system masses, for example \Msys $\simeq 1$~\Msun, the derived $\mu$ values still imply an appreciably large companion. Another issue is the observed properties of the jet. The $\mu$ values obtained with the precession model suggest a jet driven by the primary, and this cannot be reconciled with the assumption that the disk seen in the HST images is associated with the primary in the system. Therefore it seems unlikely that a precession model is an appropriate explanation for the observed wiggle.

In the orbital model, we find a mass-ratio that implies the jet-driving source is not the primary in the system. As discussed in Section~\ref{subsub:inconsistencies} this outcome is very reasonable and is supported by observational evidence. If this is an equal-mass binary, we would anticipate a different spectral type than what the observed stellar spectrum suggests. If one assumes that the primary conforms to an M0 or M1 spectral type ($M_p \simeq 0.5$~\Msun), we argue that this model gives good agreement for \Msys\, $\sim 0.6-0.7$~\Msun. Additionally, the orbital model curve seen in Figure~\ref{fig:prec_orb_fit} best matches the morphology of the jet. All of these arguments provide compelling evidence that if the curvature can be explained by a wiggling jet model, the orbital motion model is a strong candidate, and that the jet is associated with the smaller, hidden star in the system.

\begin{figure}
\centering
\includegraphics[width=\linewidth]{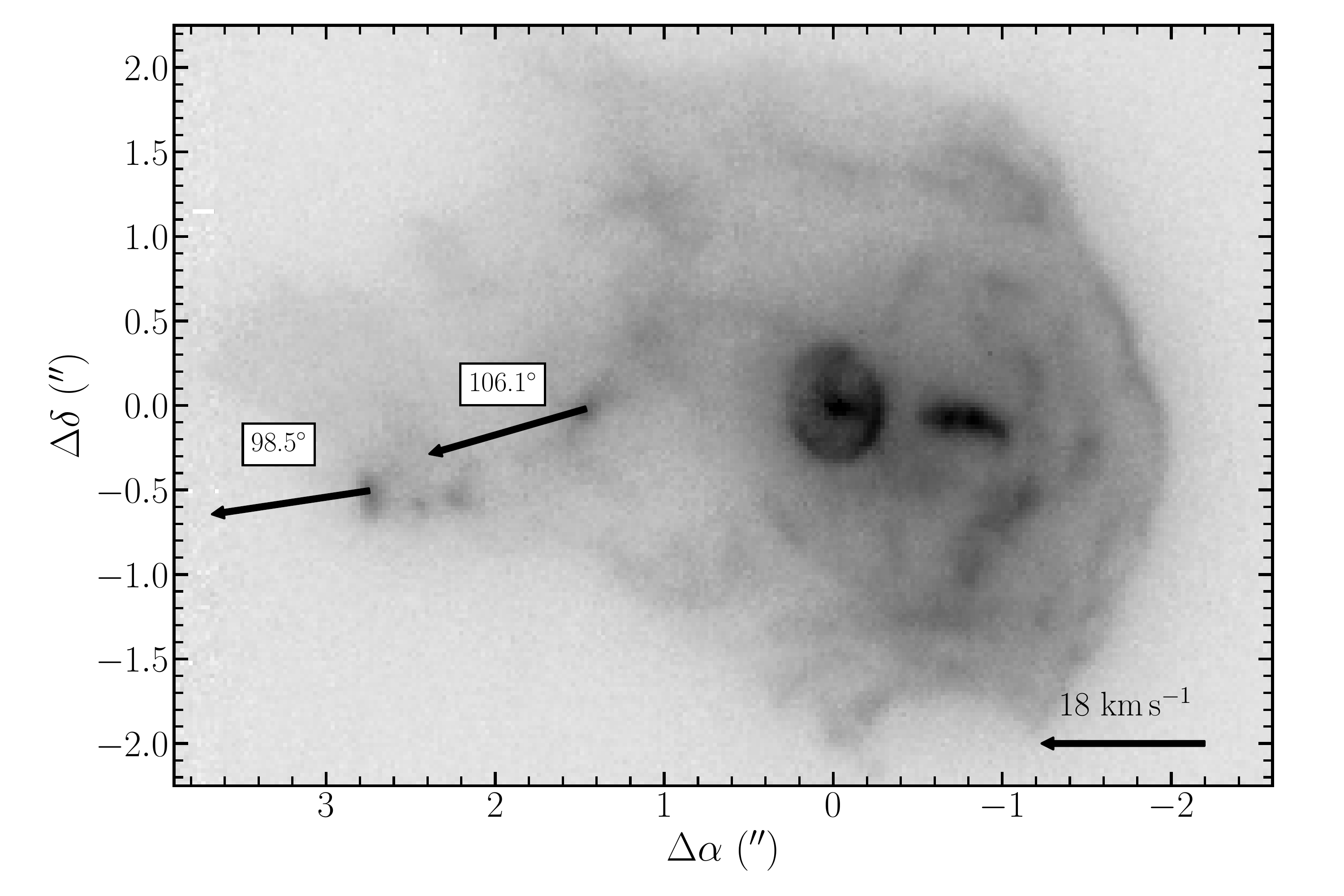}
\caption{Proper motion vectors superimposed on the \Oi\, emission line. A small difference is observed, pointing overall radially outward though possibly pointing along the direction of the jet curvature. The angles are measured east of north in the plane of the sky.}
\label{fig:prop_mot_vectors}
\end{figure}

We have not explored the impact of stellar winds here. Generally, side-wind deflection models imply a hyperbolic curvature as the jet is deflected away from the wind source~\citep{Raga2009AA}, although more complicated morphologies in photoionized regions are possible~\citep{Masciadri2001AJ}. In our observations, the C-shaped morphology anticipated by a deflection model appears to be applicable primarily to the envelope but not the jet, indicating that the jet is either shielded to some degree from the winds or that multiple winds are influencing the system in a way that is beyond the scope of this paper. 
{Furthermore, using the derived proper motions and overlaying their vectors on the data (see Figure~\ref{fig:prop_mot_vectors}), it is seen that the knots do point radially outward from the source with a slight difference} {($\sim \ang{7}$)} {in their directions. The decrease in angle exhibited by knot E1 could potentially indicate an interaction between the jet and a side-wind; however with proper motions available for only two knots either conclusion is approached with caution.}
Nevertheless these models lay a positive groundwork, as future observations may help provide greater constraints on the parameters of a potential companion.

\subsection{Diagnostics and mass-loss}
\label{subsec:discuss_diags}

Even if we could not reliably determine the temperature of the jet in our data through traditional diagnostic diagrams, the presence of refractory species like [\ion{Fe}{ii}] and [\ion{Ni}{ii}] provides opportunity to explore ranges of density, as well as the potential role of fluorescent pumping of the jet. It is reasonable to assume that the presence of external ionizing sources may induce fluorescence in the jet, particularly given the presence of the [\ion{Ni}{ii}] lines and the observed line ratios in Figure~\ref{figure:fluor_plot}. However, the density ranges estimated from the [\ion{Ni}{ii}] lines ($\sim10^4 - 10^5$~cm$^{-3}$) are above the critical limit for fluorescent pumping, which suggests that shock processes play the larger role in producing the emission lines. This may be due to the envelope shielding the jet to some degree from incident radiation, making it not as exposed as others to the external environment.

This result has interesting implications when considering the mass-loss rate. In Section~\ref{sub:mass_loss} we presented limits on $\dot{M}_j$ in the range $10^{-11}-10^{-10}$~\Msun~yr$^{-1}$ and argued that even if photoionization plays a large role, we would not expect a mass-loss greater than $10^{-9}$~\Msun~yr$^{-1}$. Additionally there is an asymmetry in the luminosities and mass-loss rates with the blueshifted jet presenting generally higher values. While the jet may benefit from shielding by the envelope, the fact that the computed values are higher in the blue lobe than the red demonstrates that the jet may still be partially exposed to external radiation. These lower values are typically seen in very low-mass protostars and brown dwarfs~\citep{Whelan2009ApJ, riaz2017}, which is consistent with the spectral typing of the source (Figure~\ref{fig:append_A1}). 

\section{Conclusions}
\label{sec:conclusions}

We have performed the first analysis of the possible origins of the curvature and physical conditions of the proplyd 244-440 jet using high spatial resolution IFU observations with MUSE NFM+AO. We identifed in the MUSE data multiple, previously unreported knots in the redshifted jet lobe. These were observed in various emission lines, most notably [\ion{O}{i}] and refractory species like [\ion{Fe}{ii}] and [\ion{Ni}{ii}]. Two knots (E1, E3) are also visible in archival HST images. 

Using measurements of the E1 and E3 knots in the MUSE and HST data, we estimated a low proper motion of $9.5 \pm 1.1$~mas yr$^{-1}$ and inferred a jet inclination angle $i_{\rm jet} = \ang{72.2} \pm \ang{4.2}$, which appears contrary to previous interpretations of a nearly edge-on disk. Closer analysis suggested that the jet is not associated with the observed disk, and we posit that the jet is actually driven by a smaller companion.

We utilized a jet-wiggling model to explore for the first time the curvature in the jet, and found that in the absence of other forces (i.e., multiple strong side-winds) the curvature could be explained by orbital motion of the jet source. As recent evidence suggests this is a spectroscopic binary, this is not unreasonable. We further reason that if this is due to a companion, we might expect the driving source to be $\leq 0.15$~\Msun\, in orbit around an M0 or M1 type star ($M \sim 0.5$~\Msun) at a separation of $\sim 30$~au.

Using the [\ion{O}{i}] and [\ion{S}{ii}] lines, we estimated a lower limit on the mass-loss rate in the jet on the order of $10^{-10} - 10^{-11}$\Msun~yr$^{-1}$.  If we assume the jet is nearly completely photoionized we set an upper limit on the mass-loss rate of $< 10^{-9}$~\Msun\, yr$^{-1}$. We note that similarly small values are observed in low-mass and substellar objects such as brown dwarfs. 

Finally, we looked at the ``proplyd lifetime problem'' and estimated an evaporation time between $0.1-0.2$~Myr. The minimum dynamical age of the jet was found to be around 300~yr, indicating that the source is still quite active and that the jet may still be quite young. Compared to the evaporation time of this proplyd, this also tells us that photoionization and photoevaporation of the proplyd had likely been occurring for some time before the jet was launch. This might indicate that the envelope has acted as a protective shell enclosing the jet and shielding it from a significant portion of Ly$\alpha$~radiation. This yields critical information about the durability of the accretion-outflow connection in the harshest of conditions. We also raise the possibility that the calculated dynamical age may drastically underestimate the true age, reasoning that if the jet has extended beyond the envelope it may have been completely destroyed. 

This work demonstrates the power of the VLT/MUSE NFM instrument in exploring jet launching dynamics. The data yielded by this instrument is rich in information, and is capable of exploring spatial structure across many emission lines critical in the study of stellar jets. The high angular resolution provided by the instrument is particularly ideal for the exploration of jets in high-radiation environments, as the external irradiation exposes quiescent, unshocked material, and allows us to better identify emission features and remove nebular and envelope contributions to the jet emission. 

\begin{acknowledgements}
    {We are grateful to the referee Alex Raga for his feedback which has helped us improve the quality of this work. }A.K. would like to acknowledge funding through the John and Pat Hume Doctoral Scholarship at Maynooth University, Ireland. C.F.M. is funded by the European Union under the European Union’s Horizon Europe Research \& Innovation Programme 101039452 (WANDA). S.F. is funded by the European Union under the European Union’s Horizon Europe Research \& Innovation Programme 101076613 (UNVEIL). Views and opinions expressed are however those of the author(s) only and do not necessarily reflect those of the European Union or the European Research Council. Neither the European Union nor the granting authority can be held responsible for them. We also extend our thanks to Monika Petr-Gotzens and Teresa Giannini for their helpful comments and discussion. 
\end{acknowledgements}

\bibliography{_ref}
\bibliographystyle{aa}

\clearpage
\begin{appendix}

\section{Supplemental Information}
\label{sec:appendixA}

\begin{figure}
\begin{center}
	\includegraphics[width=\linewidth]{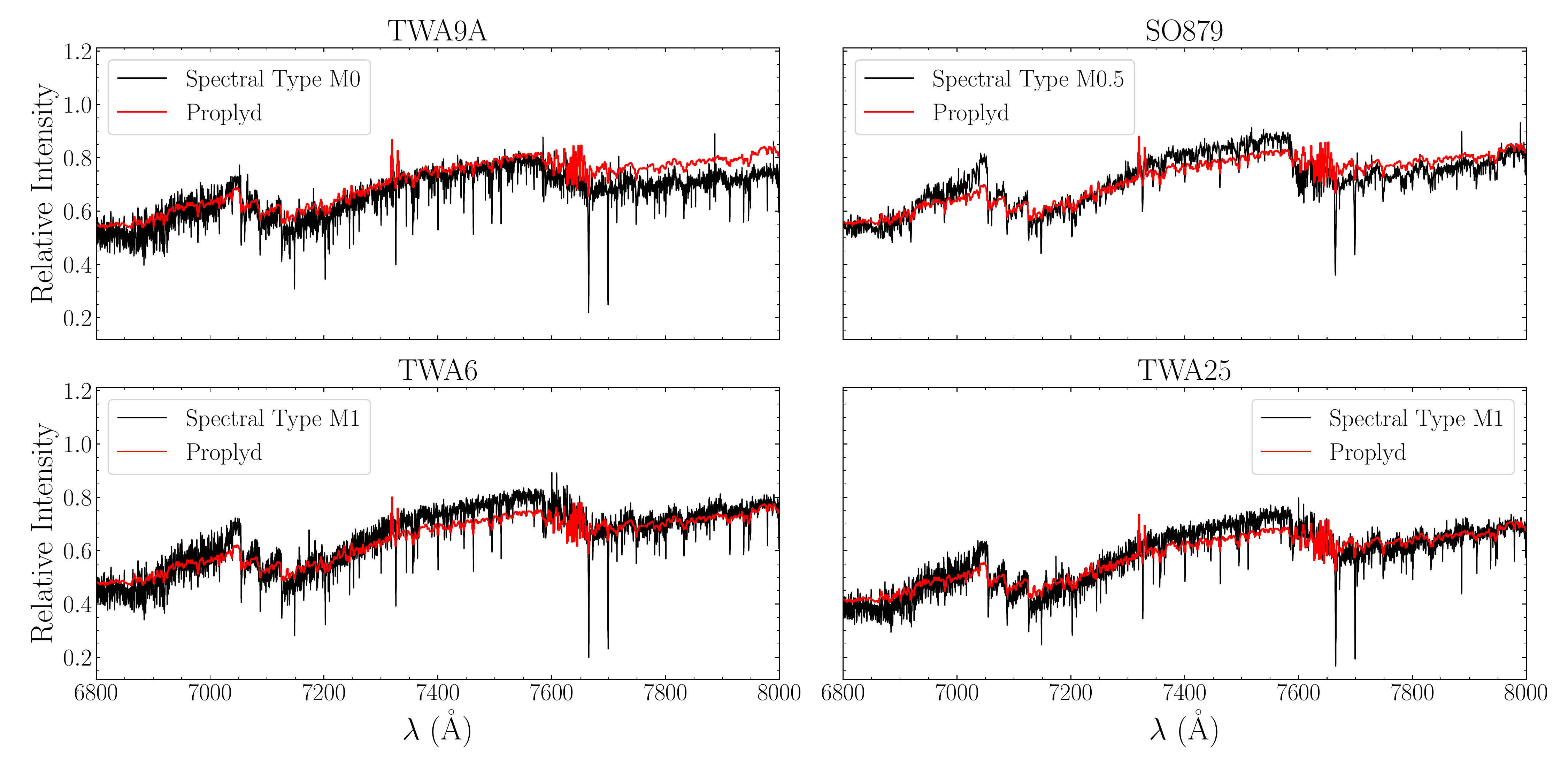}
\end{center}
\caption{On-source spectrum of 244-440 (red) overplotted with spectra from a sample of M-type YSOs. The spectral type of each YSO is given in the legend of each panel.}
\label{fig:append_A1}
\end{figure}

Many of the results discussed in this paper require us to have at least a broad estimate of the source mass. If we are to utilize a jet wiggling model, for example, a key assumption is that the source contains a binary as a critical parameters is the ratio of the companion mass to the primary. Similarly, any jet proper motion study or computation of mass-outflow rates must be compared against some mass if we are to determine how reasonable our values are. To accomplish this, we extracted an on-source spectrum, corrected for extinction, and compared it with several spectra of known stellar types to find which is most similar as shown in Figure~\ref{fig:append_A1}.

Our spectral templates were all observed on the X-Shooter instrument and their spectral types reported in~\citet{Manara2013AA} and~\citet{Manara2017AA}. Based on this, we argue that 244-440 is most likely an M0 or M1 spectral type star, placing it on the very low-mass end. We believe this to be reasonable as well, as the proper motions and radial velocities are both quite low, indicative of a low-power outflow. Additionally, the estimated mass-outflow rate is comparable with those seen in brown dwarfs, strengthening this argument~\citep{Whelan2014AA, RiazApJ2015b, riaz2017}. 

\begin{figure}
  \centering
    \includegraphics[width=0.48\linewidth, trim={0.1cm 0.1cm 0.05cm 0.05cm}, clip=true]{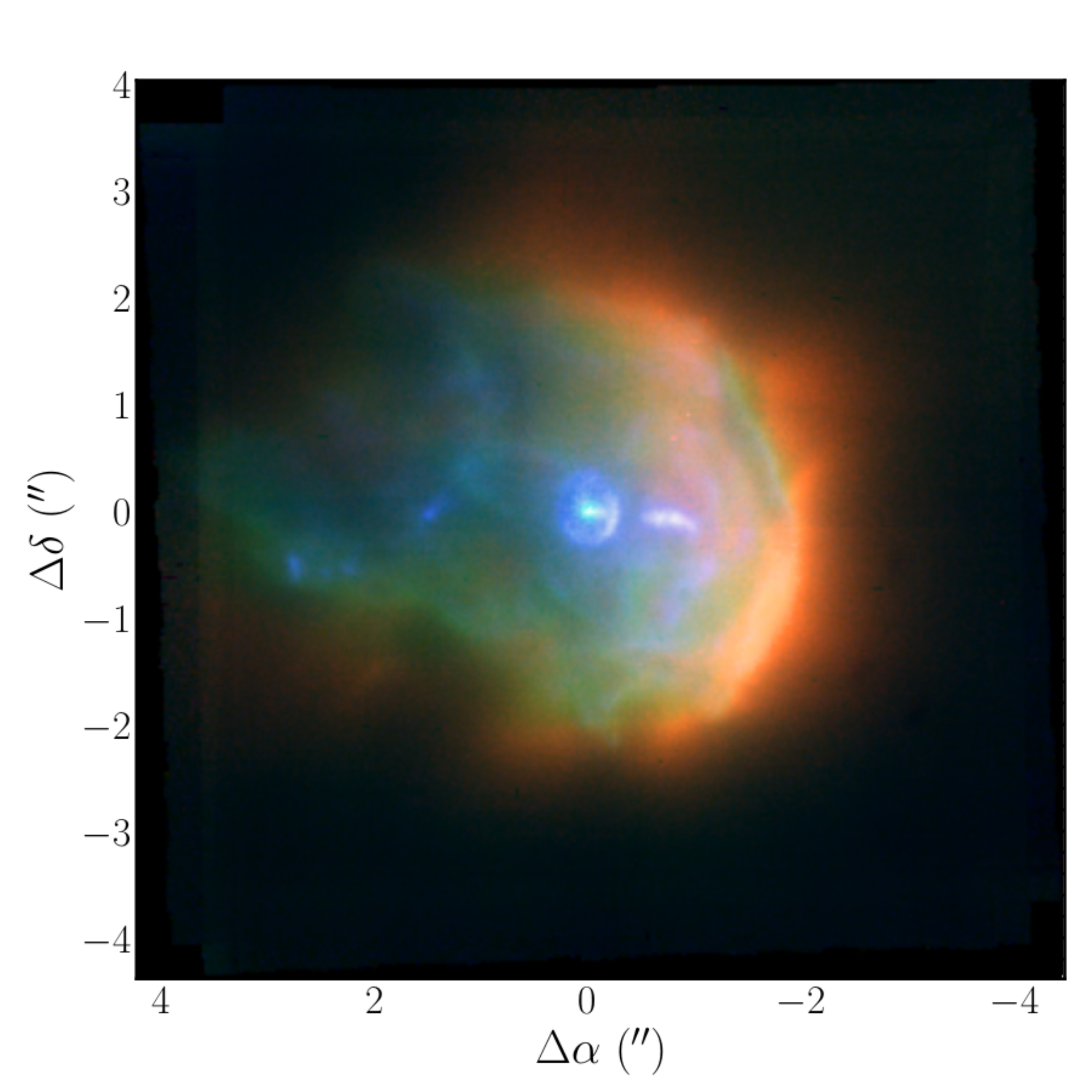}
    \includegraphics[width=0.48\linewidth, trim={0.1cm 0.1cm 0.05cm 0.05cm}, clip=true]{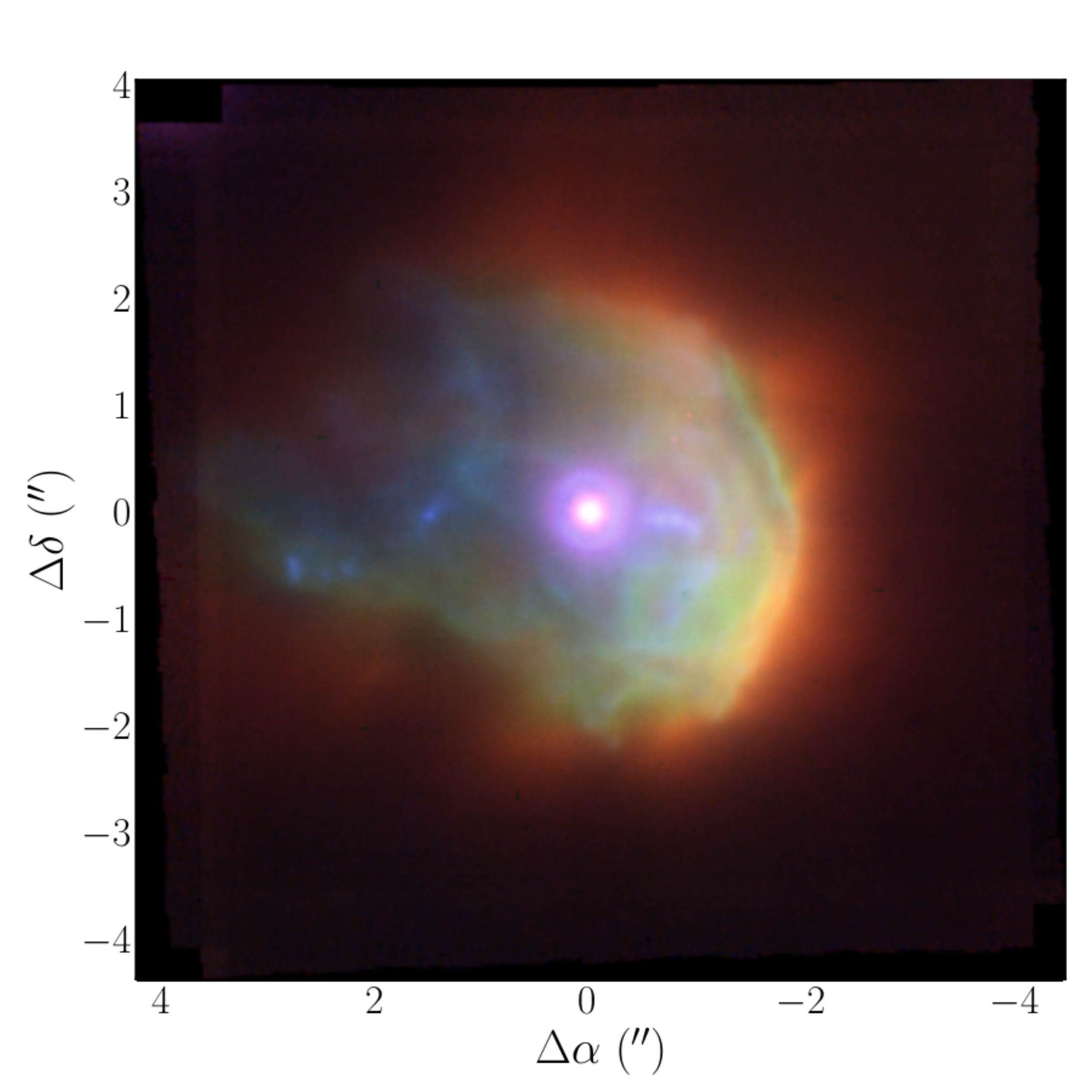}
    \caption{Three-color flux-integrated image composites of the proplyd 244-440 for the background-subtracted (left) and unsubtracted (right) data. Red is [\ion{Ar}{III}]$\lambda7136$, green is \Ha, and blue is [\ion{O}{i}]$\lambda6300$. }
    \label{fig:append_A2}
\end{figure}

\begin{figure}
  \centering
    \includegraphics[width=0.48\linewidth, trim={0.05cm 0cm 0cm 0cm}, clip=true]{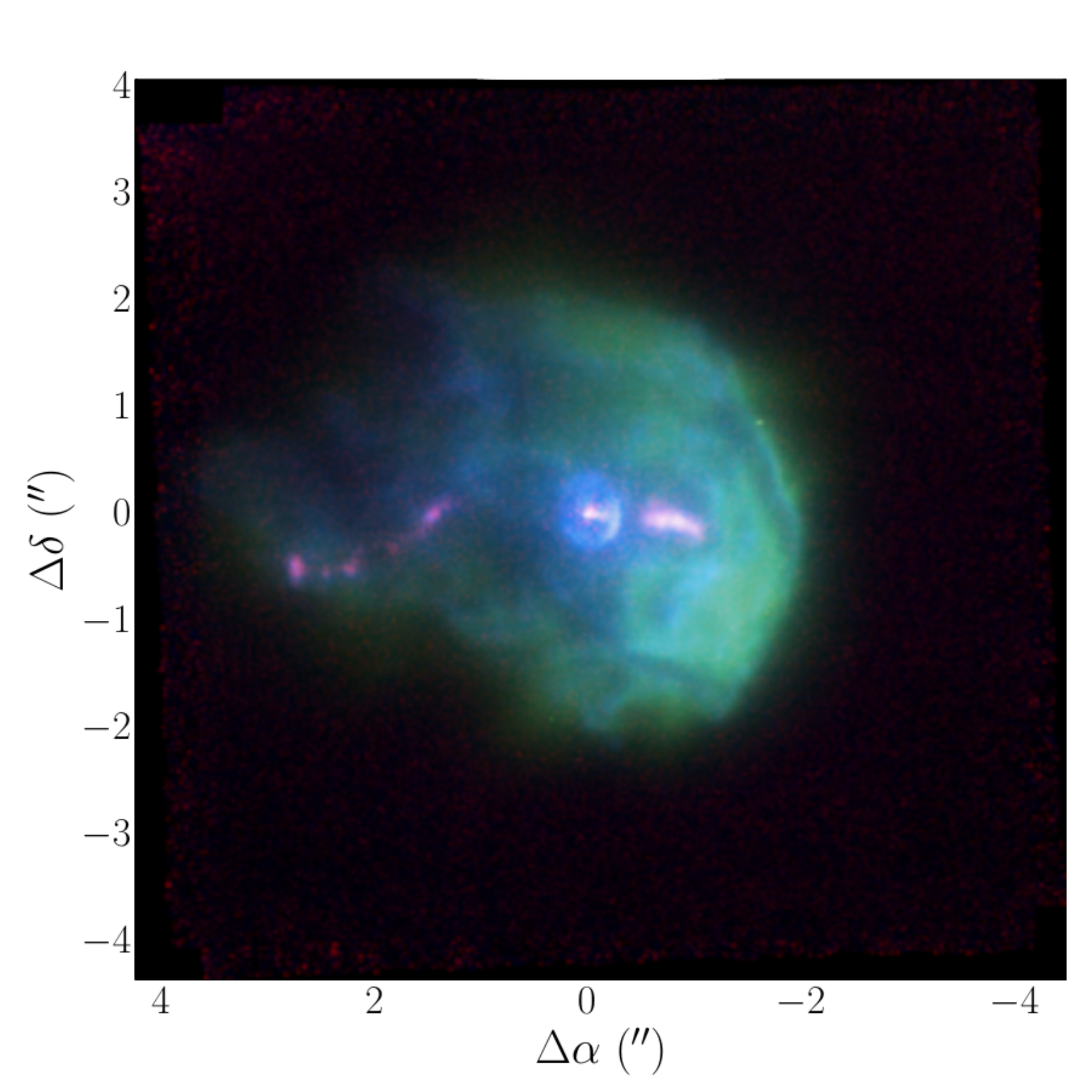}
    \includegraphics[width=0.48\linewidth, trim={0.05cm 0cm 0cm 0cm}, clip=true]{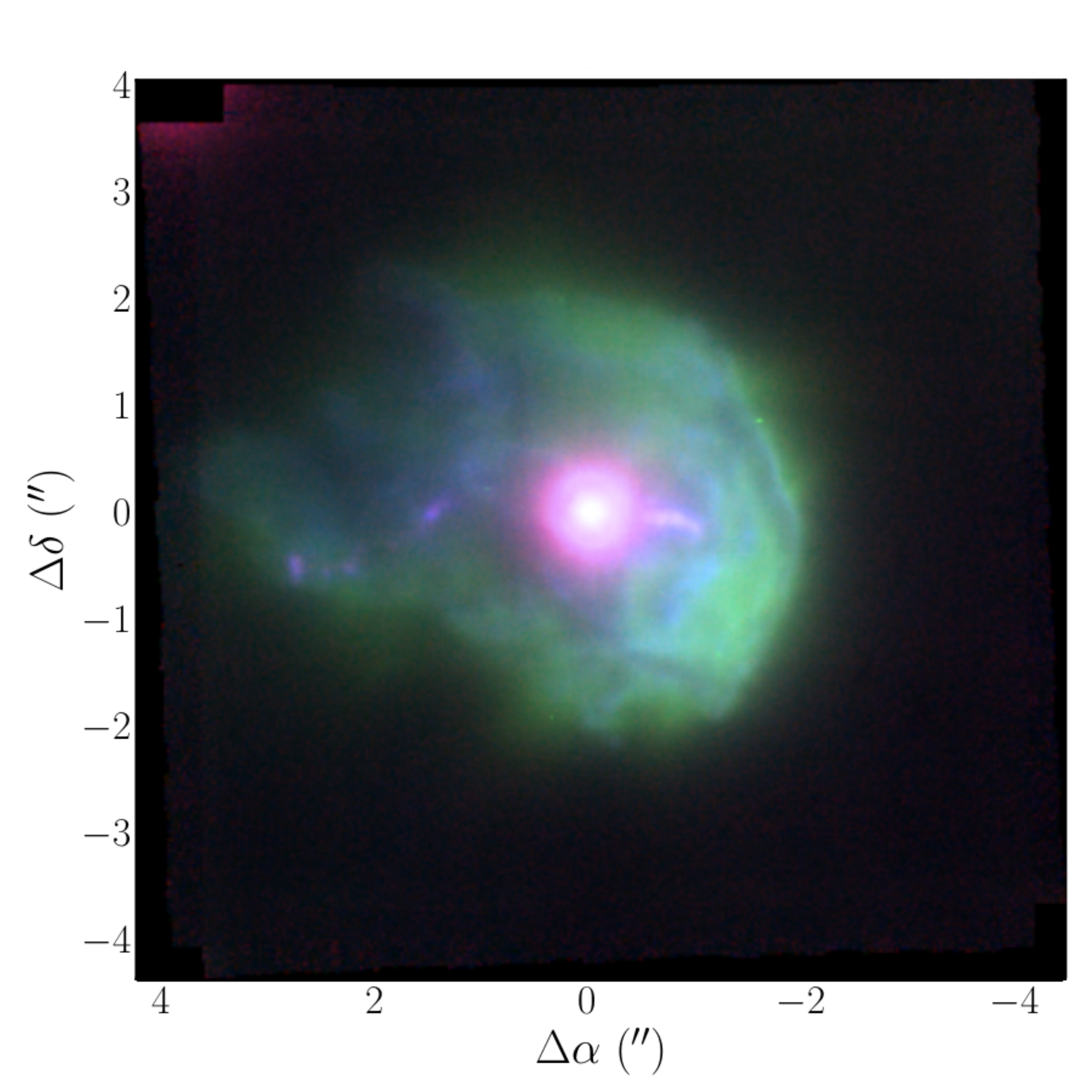}
    \caption{Same as Figure~\ref{fig:append_A2}. Red is [\ion{Fe}{ii}]$\lambda7155$, green is [\ion{N}{ii}]$\lambda6548$, and blue is [\ion{O}{i}]$\lambda6300$.}
    \label{fig:append_A3}
\end{figure}

\begin{figure}
    \centering
    \includegraphics[width=0.48\linewidth, trim={0.1cm 0.1cm 0.05cm 0.05cm}, clip=true]{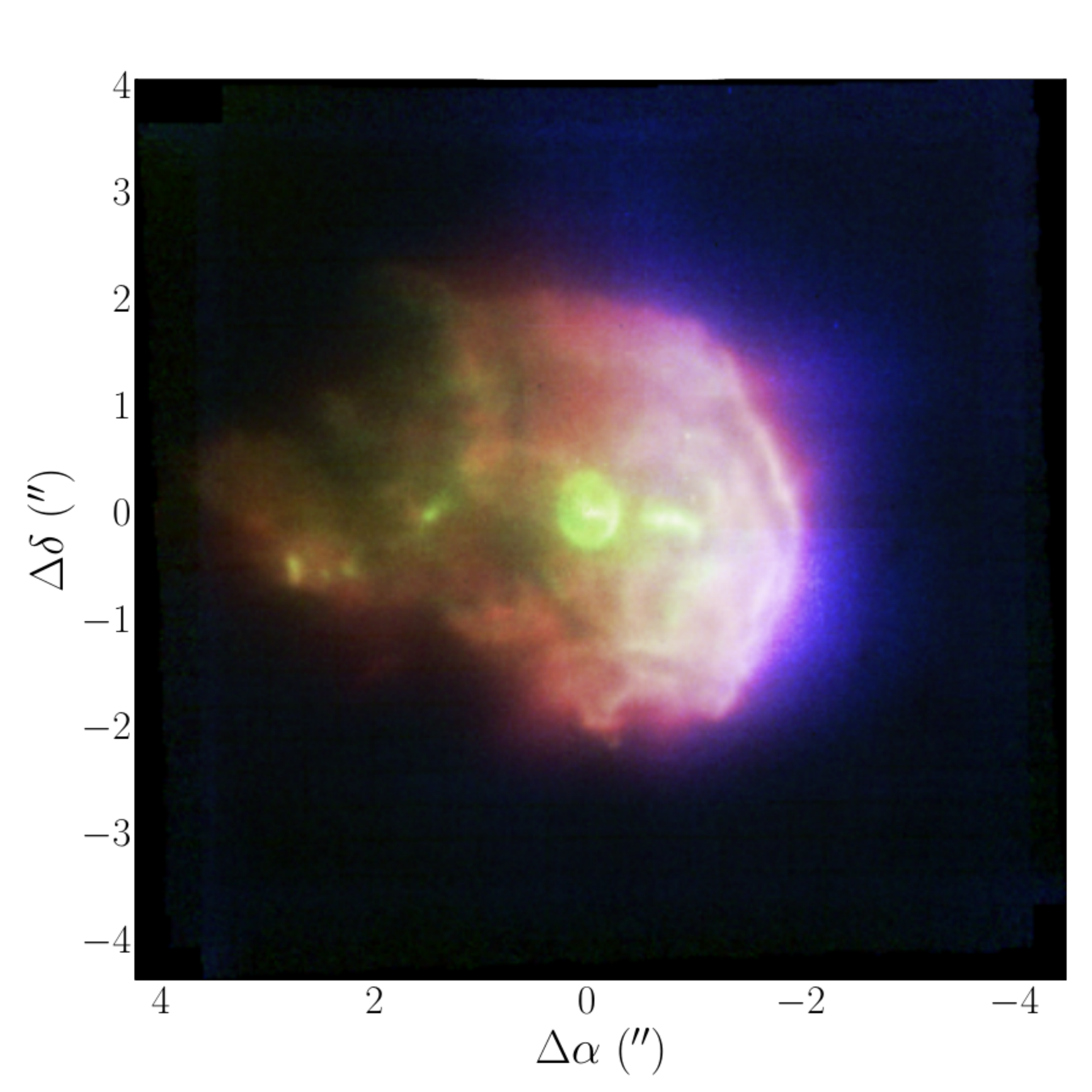}
    \includegraphics[width=0.48\linewidth, trim={0.1cm 0.1cm 0.05cm 0.05cm}, clip=true]{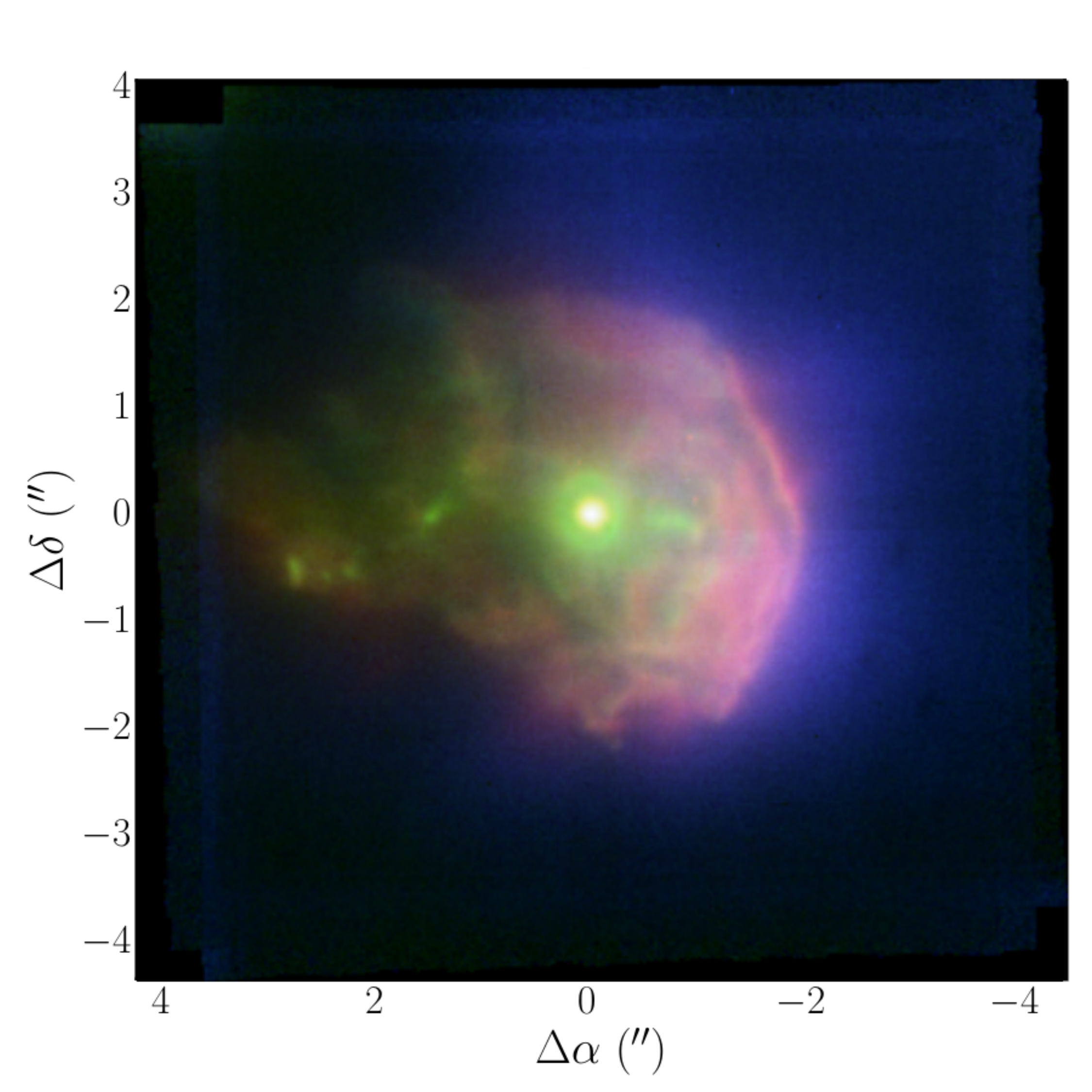}
    \caption{Same as Figure~\ref{fig:append_A2}. Red is [\ion{O}{ii}]$\lambda7320$, green is [\ion{O}{i}]$\lambda6300$, and blue is [\ion{O}{iii}]$\lambda5007$.}
    \label{fig:append_A4}
\end{figure}

\begin{figure}
    \centering
        \includegraphics[width=0.48\linewidth, trim={0cm 0cm 0cm 0cm}, clip=true]{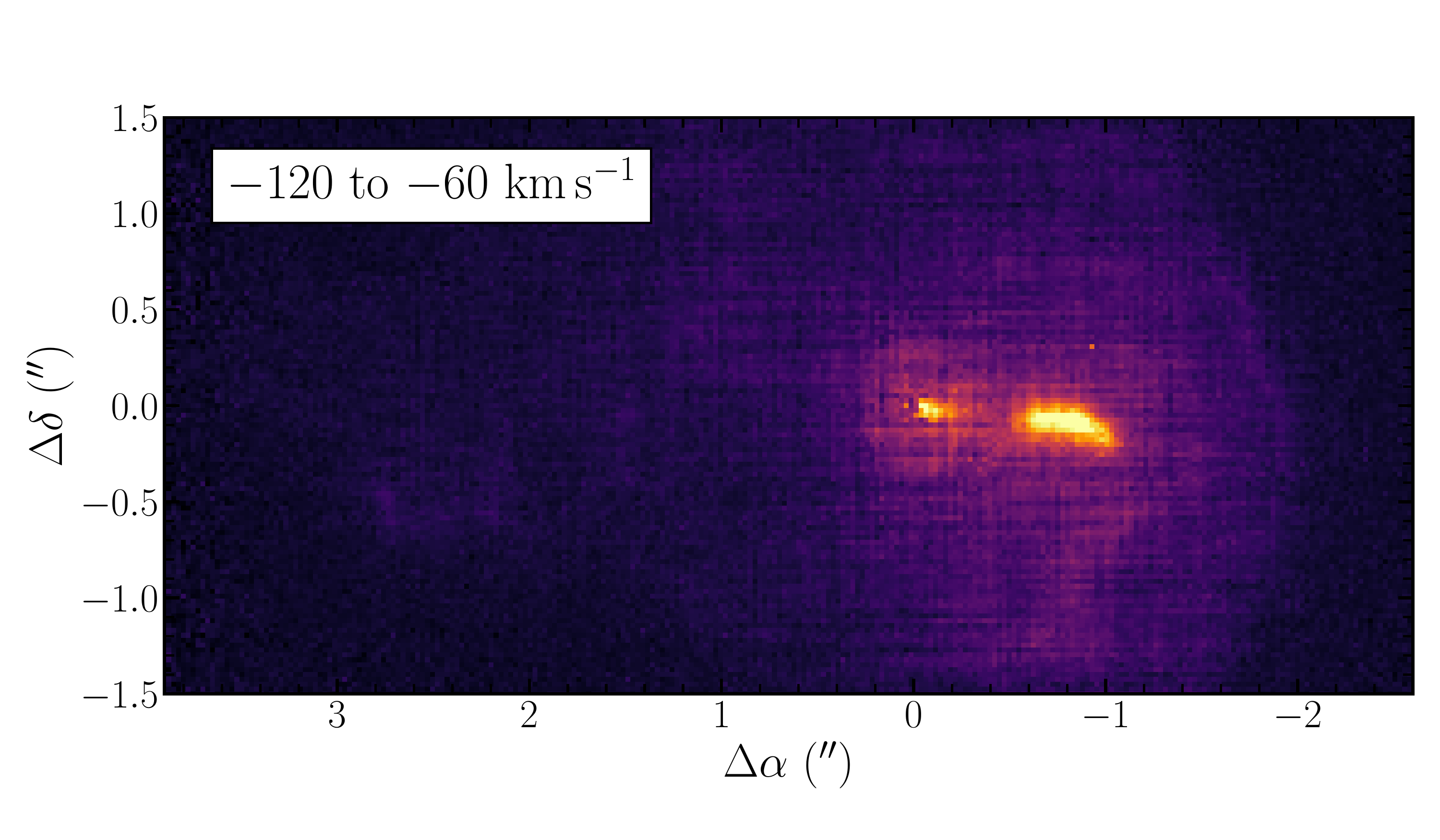}
        \includegraphics[width=0.48\linewidth, trim={0cm 0cm 0cm 0cm}, clip=true]{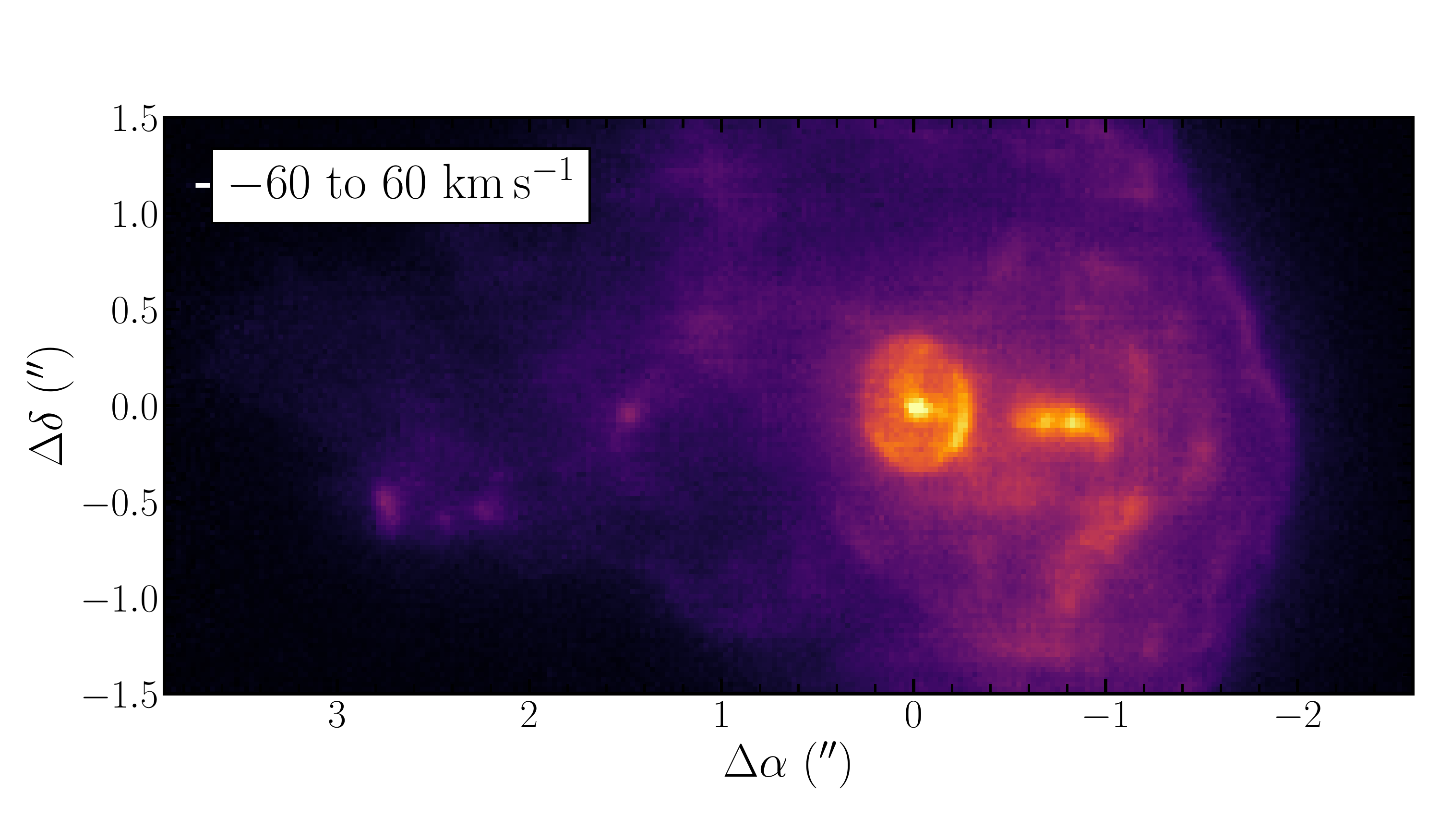}
        \includegraphics[width=0.48\linewidth, trim={0cm 0cm 0cm 0cm}, clip=true]{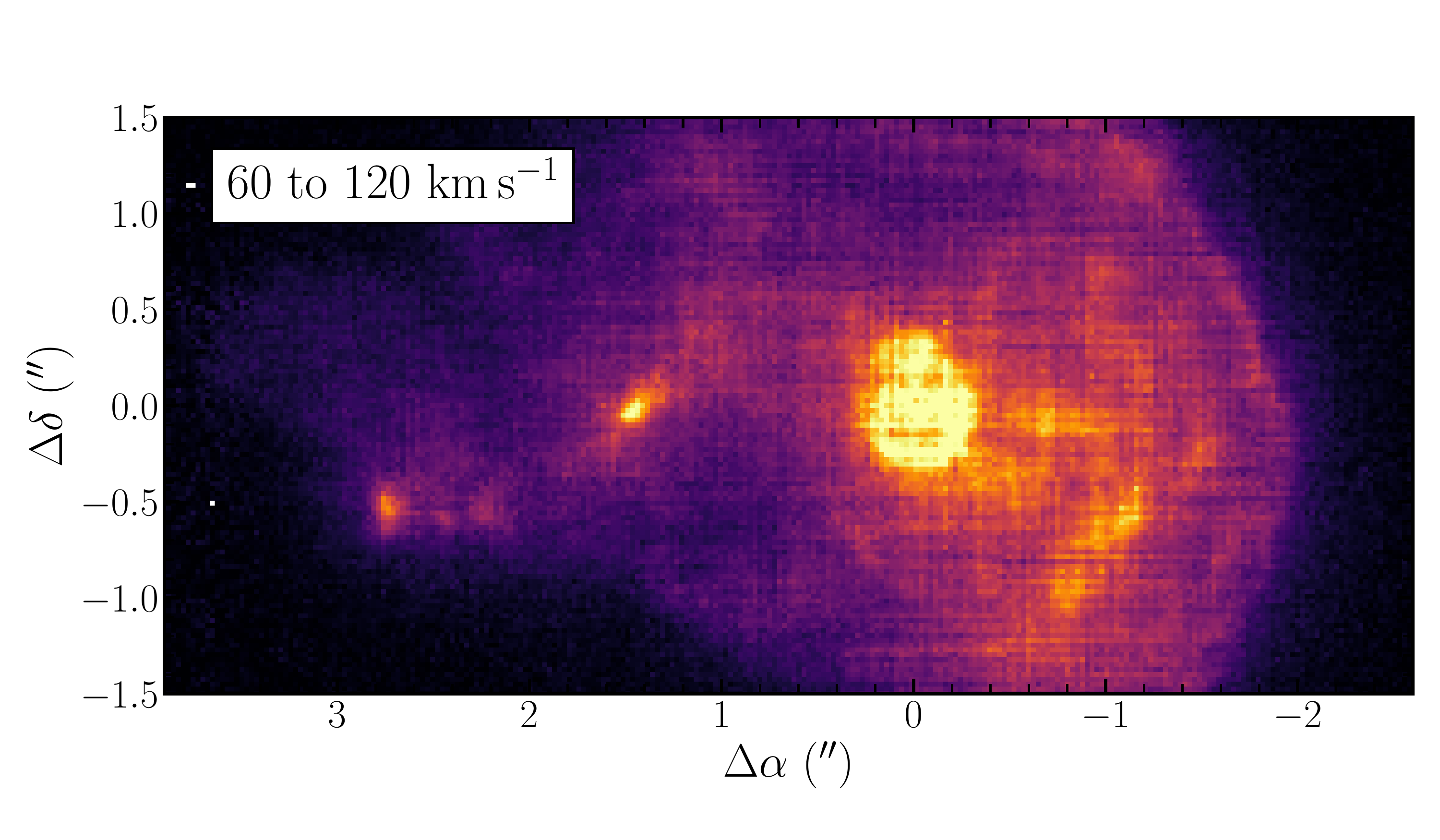}
    \caption{Velocity channel maps of the [\ion{O}{i}]$\lambda6300$ line in the MUSE data.}
    \label{fig:chanmaps01}
\end{figure}

\begin{figure}
    \centering
        \includegraphics[width=0.48\linewidth, trim={0cm 0cm 0cm 0cm}, clip=true]{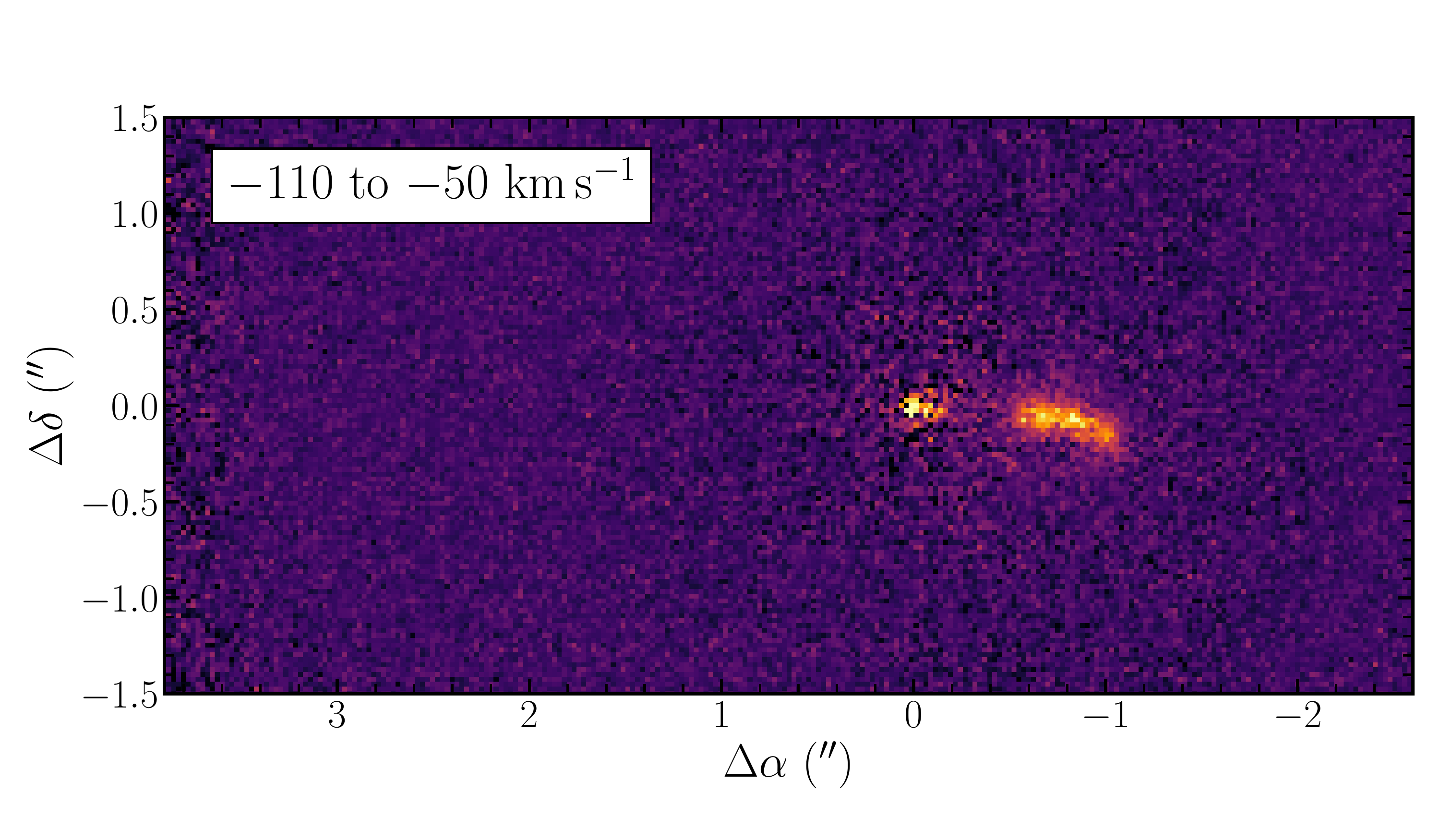}
        \includegraphics[width=0.48\linewidth, trim={0cm 0cm 0cm 0cm}, clip=true]{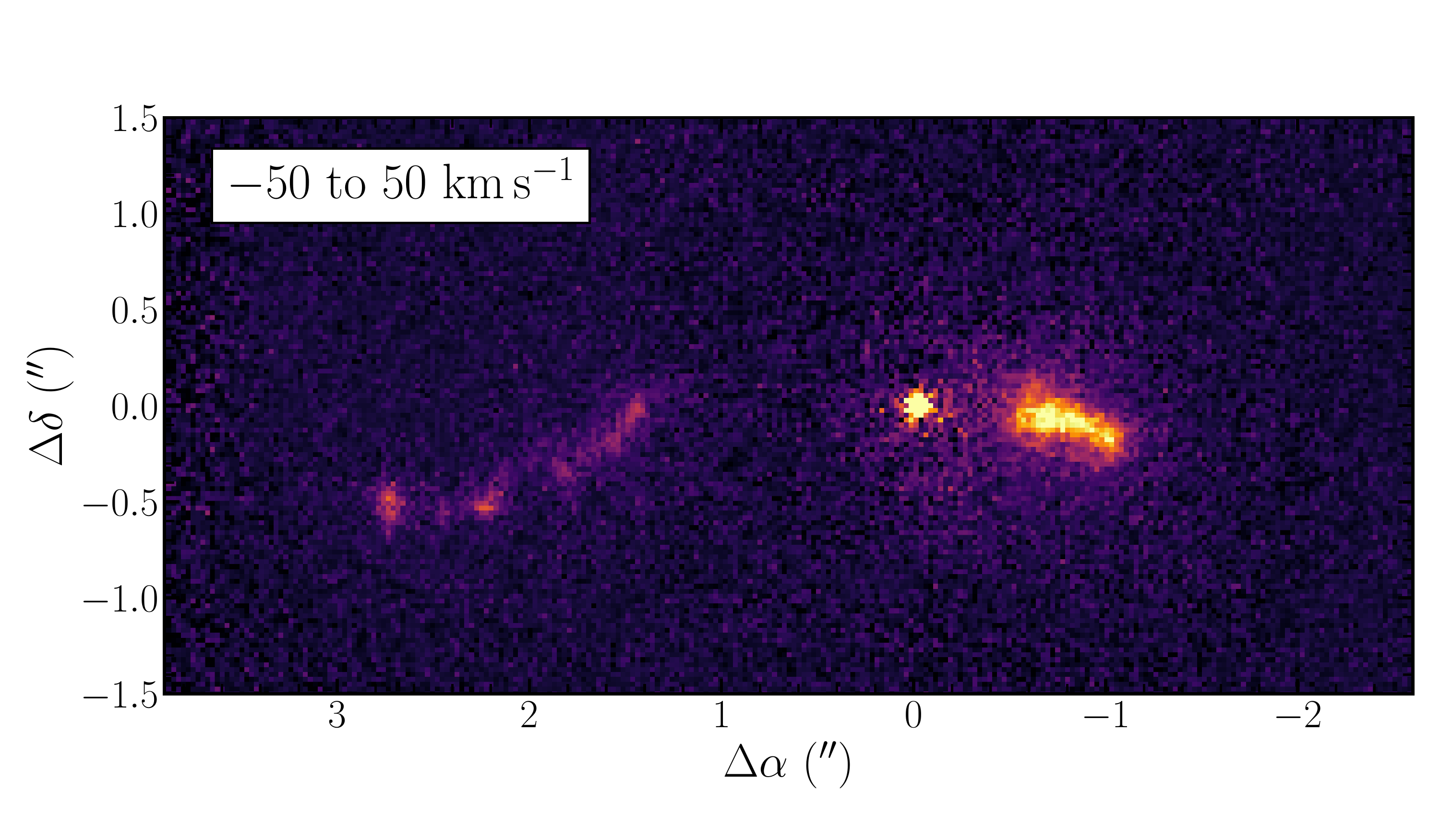}
        \includegraphics[width=0.48\linewidth, trim={0cm 0cm 0cm 0cm}, clip=true]{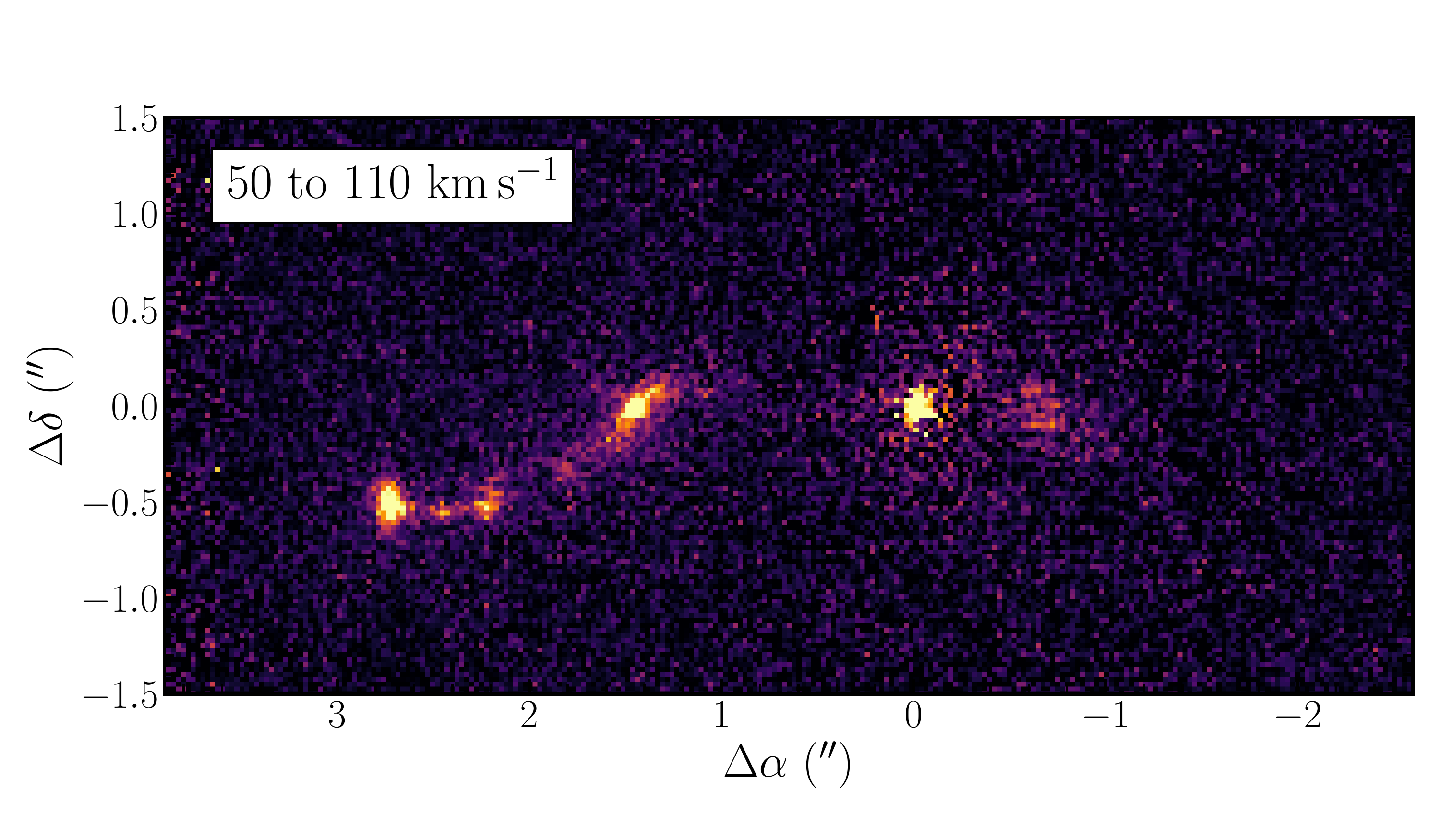}
    \caption{Same as Figure~\ref{fig:chanmaps01}, but for the [\ion{Fe}{ii}]$\lambda7155$ line.}
    \label{fig:chanmaps02}
\end{figure}

\begin{figure}
  \centering
    \includegraphics[width=0.6\linewidth, trim={0.1cm 0.05cm 0.05cm 0.05cm}, clip=true]{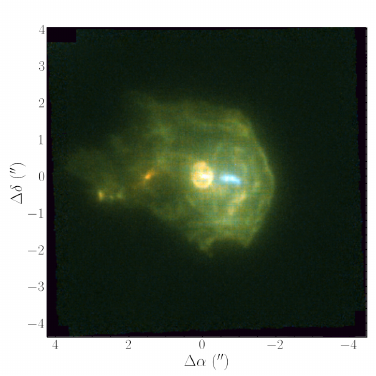}
    \caption{Three-color composite of the [\ion{O}{i}]$\lambda6300$ emission line. Green is the full flux-integrated image, and red and blue are the red- and blueshifted emissions shown in Figure~\ref{fig:chanmaps01}.}
    \label{fig:rgb_chanmap01}
\end{figure}

\begin{figure}
  \centering
    \includegraphics[width=0.6\linewidth, trim={0.1cm 0.05cm 0.05cm 0.05cm}, clip=true]{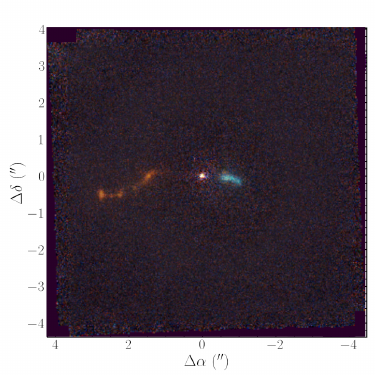}
    \caption{Three-color composite of the [\ion{Fe}{ii}]$\lambda7155$ emission line. Green is the full flux-integrated image, and red and blue are the red- and blueshifted emissions shown in Figure~\ref{fig:chanmaps02}.}
    \label{fig:rgb_chanmap02}
\end{figure}

Figures~\ref{fig:append_A2}$-$\ref{fig:append_A4} show flux-integrated three-color composites for various emission lines extracted from both background-subtracted and unsubtracted cubes.  Figures~\ref{fig:chanmaps01} and~\ref{fig:chanmaps02} show velocity channel maps isolate the redshifted and blueshifted emission features. Figures~\ref{fig:rgb_chanmap01} and~\ref{fig:rgb_chanmap02} show these features in color composites using flux-integrated spectro-images.

\end{appendix}
\end{document}